\documentclass[11pt, onecolumn]{article}

\usepackage{hyperref}
\usepackage{amsfonts}
\usepackage{amssymb}
\usepackage{amsmath}
\usepackage{bbm}

\usepackage{color}
\usepackage{mathrsfs}
\usepackage[refpage, notintoc]{nomencl} 
\usepackage[version=3]{mhchem}
\usepackage{chemarr}

\usepackage{amsfonts}
\usepackage[dvipdf]{graphicx}
\usepackage{amsmath}
\usepackage{amsthm}
\usepackage{amssymb}
\usepackage{mathtools}
\usepackage{bm}
\usepackage{algorithmic}
\usepackage{color,soul}
\usepackage{tensor}
\usepackage{hyperref}
\usepackage{bbm}
\usepackage{dsfont}
\usepackage[T2C]{fontenc}
\usepackage{hyperref}
\usepackage{float}
\usepackage[caption=false]{subfig}
\usepackage{multirow}
\usepackage{epstopdf}
\usepackage{array}
\usepackage{IEEEtrantools}
\usepackage{sidecap}
\usepackage{cleveref}

\newcommand{\D}{\,d}
\newcommand{\R}[1]{\mathrm{#1}}
\newcommand{\mean}[1]{\mathbb{E}\left[#1\right]}

\newcommand{\kn}{k_{\R{on}}}
\newcommand{\kf}{k_{\R{off}}}
\definecolor{dartmouthgreen}{rgb}{0.05, 0.5, 0.06}

\newcommand*{\defeq}{\mathrel{\vcenter{\baselineskip0.5ex \lineskiplimit0pt
                     \hbox{\scriptsize.}\hbox{\scriptsize.}}}%
                     =}
\newcommand*{\defeqrev}{=\mathrel{\vcenter{\baselineskip0.5ex \lineskiplimit0pt
                     \hbox{\scriptsize.}\hbox{\scriptsize.}}}}

\graphicspath{ {figures/} }

\title{\textbf{Stochastic models of gene transcription with upstream drives: \\
Exact solution and sample path characterisation}}

\author{Justine Dattani and Mauricio Barahona\footnote{m.barahona@imperial.ac.uk}\\ Department of Mathematics \\ Imperial College London \\ London SW7 2AZ, United
  Kingdom}

\date{}

\begin{document}

\maketitle

\begin{abstract}
Gene transcription is a highly stochastic and dynamic process. As a result, the 
mRNA copy number of a given gene is heterogeneous
both between cells and across time. 
We present a framework to model gene transcription in populations of 
cells with time-varying (stochastic or deterministic) 
transcription and degradation rates. Such rates can be 
understood as upstream cellular drives representing the effect of
different aspects of the cellular environment.
We show that the full solution of the master equation contains two components:
a model-specific, upstream effective drive, which encapsulates the 
effect of cellular drives (e.g., entrainment, periodicity or promoter randomness), 
and a downstream transcriptional Poissonian part,
which is common to all models.
Our analytical framework treats cell-to-cell and dynamic variability
consistently, unifying several approaches in the literature. 
We apply the obtained solution to characterise different models of experimental relevance, and to
explain the influence on gene transcription of synchrony, stationarity, ergodicity, as well as the
effect of time-scales and other dynamic characteristics of drives.
We also show how the solution can be applied to the analysis of noise sources in single-cell data,
and to reduce the computational cost of stochastic simulations.
\end{abstract}

\maketitle

\section{Introduction}
Gene transcription, the cellular mechanism through which DNA
is copied into mRNA transcripts, 
is a complex, stochastic process involving small numbers of molecules~\cite{Levine:2003}.
As a result, the number of mRNA copies for most genes is 
highly heterogeneous over time within each cell,
and across cells in a population~\cite{Raj:2008, Owens:2016, Singer:2014}.
Such fundamental randomness is biologically relevant: it underpins the 
cell-to-cell variability linked with phenotypic outcomes and cell 
decisions~\cite{Maamar:2007, 
Chang:2008, Balazsi:2011, Garcia-Ojalvo:2012, Wolf:2015}.

The full mathematical analysis of gene expression variability requires
the solution of \emph{master equations}. 
Given a gene transcription model, 
its master equation (ME) is a differential-difference equation
that describes the evolution of $P(n,t)$, the probability of 
having $n$ mRNA molecules in a single cell at time $t$.
However, MEs are problematic to solve, both analytically 
and numerically, due to the difficulties associated with discrete stochastic 
variables---the molecule number $n$ is an integer~\cite{Wilkinson:2009}.
Most existing analytical solutions of the ME are 
specific to particular models and are typically obtained via the probability generating 
function under stationarity assumptions~\cite{Hornos:2005, Visco:2008,  
Zhang:2012, Pendar:2013, Kumar:2014}. A few other solutions include the decaying time-dependence describing the relaxation transient towards stationarity from a given 
initial distribution~\cite{Iyer-Biswas:2009, Shahrezaei:2008, Vandecan:2013, Smith:2015}.  
In the usual situation when analytical solutions are intractable, 
the first few moments of the distribution are approximated, usually at stationarity, 
although error bounds are difficult to obtain under closure schemes~\cite{Paulsson:2005, Sanchez:2013}.
Alternatively, full stochastic simulations are used, yet the computational cost to 
sample $P(n,t)$ at each $t$ is often impractical, 
and many methods lead to estimation bias in practice~\cite{Zeron}.

The emergence of accurate time-course measurements of 
mRNA counts in single cells~\cite{Owens:2016, Crosetto:2015, 
Park:2014, Singer:2014, Young:2012} has revealed the high dynamic variability of 
gene expression both at the single-cell and population levels.  
This variability has several sources. Cells express genes heterogeneously~\cite{Junker:2014, 
Singer:2014} and hence models need to capture intercellular variability; but cells are also subjected to time-varying inputs of a stochastic and/or deterministic nature, either from their environment or from regulatory gene networks inside the cell.
Therefore, standard ME models with stationary solutions, which also tacitly assume that gene expression is uncorrelated between cells, cannot capture fully such sources of variability.
Mathematically, ME models must be able to 
describe time-dependent gene transcription in single cells within an inhomogeneous population, 
i.e., they must allow a varying degree of synchrony and of cell-to-cell variability across the population.
In addition, they must be able to account for non-stationary dynamic variability 
due to upstream biological drives, such as circadian rhythms and 
cell cycle~\cite{Mihalcescu:2004, Bieler:2014}, 
external signalling~\cite{Corrigan:2014}, or 
stimulus-induced modulation or 
entrainment~\cite{Molina:2013, Konermann:2013}. 

Recent techniques to model cell-to-cell correlations have used 
the marginalisation of extrinsic components~\cite{Zechner:2014}, 
mixed-effects models~\cite{Almquist:2015}, or deterministic rate 
parameters~\cite{Jahnke:2007}. Several of the results correspond to deterministic 
rates and are well-known in queuing theory~\cite{Eick:1993}.  
However, full solutions of the ME that capture temporal heterogeneity as well as 
variability in parameters, from the 
single-cell to the population level, are yet 
to be explored, and could help unravel conjunctly with experiments 
how the dynamics of upstream drives 
within a biological network affect gene transcription.

Here, we consider a simple, yet generic, framework for the solution of 
the ME of gene transcription and degradation for single cells 
under upstream drives, i.e., when the transcription and 
degradation parameters are time-dependent functions or stochastic variables.
We show that the exact solution $P(n,t)$ for such a model naturally 
decouples into two parts: a discrete transcriptional Poissonian downstream component, 
which is common to all transcription models of this kind, 
and a model-specific continuous component, 
which describes the dynamics of the parameters reflecting the upstream variation. 
To obtain the full solution $P(n,t)$ one only needs to calculate the 
probability density for the model-specific upstream drive, 
which we show corresponds to a continuous variable satisfying 
a linear random differential equation directly related to traditional 
differential rate equations of chemical kinetics. 
Our results can thus be thought of as a generalisation 
of the Poisson representation~\cite{Gardiner:1977,Gardiner}
(originally defined as an \textit{ansatz} with constant rate parameters) 
to allow for both time-varying and stochastic rates  
in transcription-degradation systems. 
Our work also departs from the work of Jahnke and Huisinga~\cite{Jahnke:2007} 
by allowing the presence of cell-to-cell variability (or uncertainty) in the dynamical drive.

Below, we present the full properties of the general solution, and we derive
the relationship of the observable time-varying moments with the moments of 
the dynamic upstream component. 
Because our framework treats dynamic and population variability consistently, we 
clarify the different effects of 
variability in the drives by considering the Fano factor across the population and across time.
To illustrate the utility of our approach, we present analytical and numerical analyses 
of several models from the literature, which are shown to simply correspond to 
different upstream drives, deterministic or stochastic. These examples highlight our 
modelling approach: a flexible solvable model with upstream dynamic
variability, reflecting the generic hypothesis that 
experimentally observed outputs are usually driven by fluctuating, 
usually unmeasurable and uncertain, upstream intra- and extra-cellular signals. 
Our framework provides a means to characterise such upstream variability, dynamical and population-wide,
and we provide examples of its use for computational biology and data analysis in relation to experiments.

\begin{figure}[htb!]
\centerline{\includegraphics[width=.54 \textwidth]{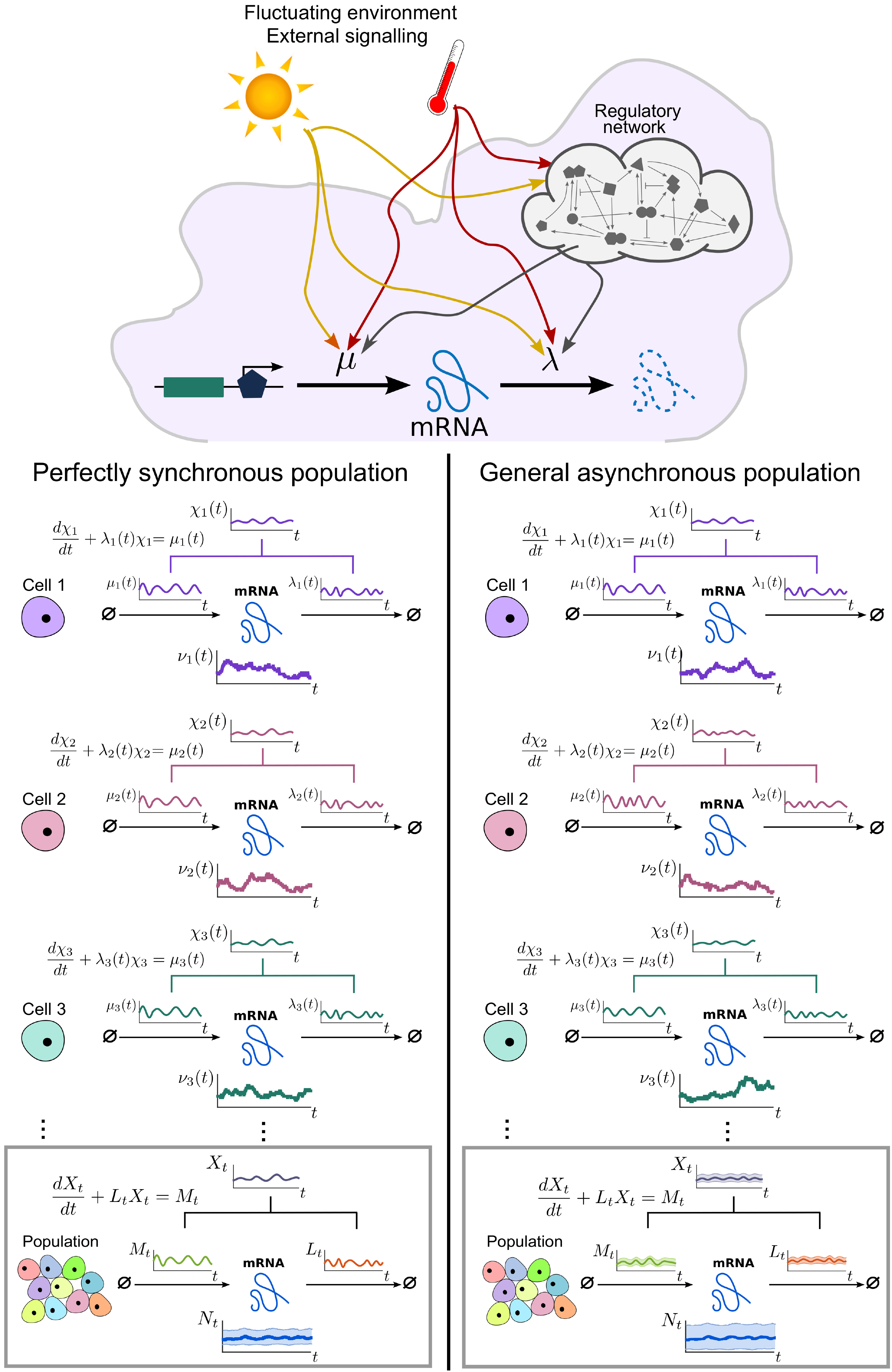}}
  \caption{Single-cell gene transcription under upstream drives. The transcription
  of each cell $i$ takes place under particular cellular drives $\{\mu_i(t)\}_{t\ge0}$ 
  and $\{\lambda_i(t)\}_{t\ge0}$, representing time-varying transcription and
  degradation rates. Both cellular drives are combined into the upstream effective drive 
  $\{\chi_i(t)\}_{t\ge0}$, which dictates the long-term probability distribution 
  describing the stochastic gene expression 
  $\{\nu_i(t)\}_{t\ge0}$ within each cell~\eqref{eq:PoiResult}.
  When there is cell-to-cell variability in the population, 
  the cellular drives are described by processes 
  $M$ and $L$ leading to the upstream effective drive $X$.
  The probability distribution of the population corresponds 
  to the mixture of the upstream process $X$ and 
  the Poissonian downstream transcriptional component, as given by~\eqref{eq:PoiMixResult}. 
Increased synchrony in the population implies decreased ensemble variability of the 
random variables $M_t$, $L_t$, $X_t$, and $N_t$.    
}
  \label{fig:summary} 
\end{figure}

\section{The master equation for gene transcription in populations of 
cells with upstream drives}

\subsection*{Notation and formulation of the problem}

To study gene expression
in a single cell with time-dependent upstream drives, 
we consider the stochastic process in continuous time $t$, 
$N=\{N_t \in \mathbb{N}: \, t\geq 0\}$, where $N_t$ is a discrete random variable describing 
the number of mRNA molecules in the cell. We look to obtain the probability mass 
function, $P(n,t) \defeq \R{Pr}(N_t=n).$

The mRNA copy number increases via transcription events 
and decreases via degradation events but, importantly, we 
acknowledge that the observed gene reflects the dynamic variability
of intra- and extra-cellular processes 
and that cells are heterogeneous.
Thus, the transcription and degradation rates can depend on time and 
can be different for each cell (Fig.~\ref{fig:summary}). 
To account for such variability, 
we describe transcription and degradation rates as stochastic processes 
$M=\{M_t \in \mathbb{R^+}: \, t\geq 
0\}$ and $L=\{L_t \in \mathbb{R^+}: \, t\geq 0\}$,
without specifying any functional form except requiring that $M$ and $L$ do not depend 
on the number of mRNA molecules already present. 
Deterministic time-varying
transcription/degradation rates, with or without cell-to-cell correlations, 
are a particular case of this definition.

Following standard notation in the stochastic processes literature, 
$M_t$ and $L_t$ denote the random variables at time $t$. 
To simplify notation, however, we depart from standard notation 
and denote the sample paths (i.e., realisations)  of $M$ and $L$ 
by $\{\mu(t)\}_{t\ge0}$ and $\{\lambda(t)\}_{t\ge0}$,
respectively, thinking of them as particular functions of time 
describing the transcription and degradation rates under 
the changing cellular state and environmental conditions 
in an `example' cell (Fig.~\ref{fig:summary}).
The sample paths of other random variables 
are denoted similarly, e.g., the sample
paths of $N_t$ are $\{\nu(t)\}_{t\ge0}$.

The sample paths $\{\mu(t)\}_{t\ge0}$ and $\{\lambda(t)\}_{t\ge0}$ represent
\emph{cellular drives} encapsulating the variability across time and across 
the population consistently.
This formulation unifies several 
models in the literature, which implicitly or explicitly assume time-varying transcription and/or 
degradation processes~\cite{Peccoud:1995, Raj:2006, Iyer-Biswas:2009, 
Sanchez:2011, Singer:2014, Senecal:2014, Zoller:2015}, and 
can be shown to correspond to particular types of dynamic 
upstream variability.
In addition, the framework allows us to specify cell-to-cell correlations 
across the population, which we refer to as the 
`degree of synchrony'. 
A population will be \emph{perfectly synchronous} when the sample paths of the 
drives for every cell in the population are identical,
i.e., if $M_t$ and $L_t$ have zero variance.
If, however, transcription and/or degradation rates
differ between cells, $M_t$ and $L_t$ themselves emerge from a probability 
density: the wider the density, the more asynchronous the cellular drives are 
(Fig.~\ref{fig:summary}).

Our aim is to obtain the probability distribution of the copy number $N_t$ under 
upstream time-varying cellular drives $M_t$ and $L_t$, themselves containing 
stochastic parameters reflecting cell-to-cell variability. We proceed in two
steps: first, we obtain the solution for the perfectly synchronous system without cell-to-cell 
variability;
then we consider the general asynchronous case.

\subsection{Perfectly synchronous population}
\label{sec:sync_deriv}

As a first step to the solution of the general case, 
consider a population of cells with perfectly synchronous 
transcription and degradation rate functions, $M=\{\mu(t)\}_{t\ge 0}$ and 
$L=\{\lambda(t)\}_{t\ge 0}$; i.e.,  
the transcription and degradation processes are defined by the same 
sample path for the whole population and the stochastic processes 
$M$ and $L$ have zero variance at all times 
(Fig.~\ref{fig:summary}).

In the perfectly synchronous case, we have an immigration-death process with reaction diagram:
\begin{align}
\label{reaction_diagram}
 \emptyset \xrightarrow{\mu(t)} \text{mRNA} \xrightarrow{\lambda(t)} \emptyset,
\end{align}
and its ME is standard:
\begin{align}
  \label{eq:ME}
& 
\frac{d}{dt}P\left(n,t|
\{\mu(\tau)\}_{\tau\in[0,t]},\{\lambda(\tau)\}_{\tau\in[0,t]}\right)  = \mu(t)P(n-1,t |
 \{\mu(\tau)\}_{\tau\in[0,t]},\{\lambda(\tau)\}_{\tau\in[0,t]}) \nonumber \\
 & \qquad + (n+1)\lambda(t) \, P(n+1,t|
 \{\mu(\tau)\}_{\tau\in[0,t]},\{\lambda(\tau)\}_{\tau\in[0,t]})  - \left(\mu(t) + n\lambda(t)\right) \, P(n,t|
  \{\mu(\tau)\}_{\tau\in[0,t]},\{\lambda(\tau)\}_{\tau\in[0,t]}),
\end{align}
where $P(n,t |\{\mu(\tau)\}_{\tau\in[0,t]},\{\lambda(\tau)\}_{\tau\in[0,t]})$ 
denotes the probability of having $n$ mRNAs at time $t$ for the 
given history of the cellular drives 
$\{\mu(\tau)\}_{\tau\in[0,t]}$ and $\{\lambda(\tau)\}_{\tau\in[0,t]}$.

Using the probability generating function 
$$G(z,t) = \sum_{n=0}^{\infty} z^n P\left(n,t|
\{\mu(\tau)\}_{\tau\in[0,t]},\{\lambda(\tau)\}_{\tau\in[0,t]}\right),$$
we transform the ME~\eqref{eq:ME} into
\begin{equation*}
  \frac{\partial G}{\partial t} = (z-1) \, \mu(t) \, G - (z-1) \, \lambda(t) \, \frac{\partial G}{\partial z}. 
\end{equation*}
Without loss of generality, let us first consider an initial condition with 
$n_0$ mRNA molecules. Using the method of characteristics, we obtain the solution:
\begin{align}
 \label{eq:G_soln}
  G(z, t \vert n_0) &= \left[(z-1)e^{-\int_0^t \lambda(\tau) \,d\tau} + 1
\right]^{n_0} e^{\chi(t)(z-1)}, 
\end{align}
which is given in terms of
\begin{align}
 \label{eq:x}
  \chi(t) \defeq \int_0^t \mu(\tau) \, e^{-\int_{\tau}^t\lambda(\tau')\,d\tau'}\,d\tau.
\end{align}
We will refer to the time-varying continuous function $\chi(t)$ as the 
\emph{effective drive}, as it integrates the effect of both cellular drives.

Notice that the solution~\eqref{eq:G_soln} can be rewritten as
the product of two probability generating functions:
\begin{align*}
 G(z, t \vert n_0) & \defeqrev G_{\R{Bin}}(z,t | n_0) \,G_{\R{Poi}}(z,t),
\end{align*}
corresponding to a binomial and a Poisson distribution, respectively. 
Hence, for the perfectly synchronous case, the solution is given by:
\begin{align}
\label{eq:NicNs}
 N_t&=N_t^{ic}|n_0+N_t^s, \quad \text{with} \\
 N_t^{ic}|n_0&\sim\R{Bin}\left(n_0,e^{-\int_0^t \lambda(\tau)d\tau}\right) \\
 \label{eq:Ns}
 N_t^s&\sim\R{Poi}\left(\chi(t)\right),
\end{align}
where $N_t^{ic}|n_0$ is a binomial random variable with $n_0$ 
trials and success probability $e^{-\int_0^t \lambda(\tau) \,d\tau}$, and 
$N_t^s$ is a Poisson random variable  with parameter $\chi(t)$. 
The physical interpretation of this breakdown is that $N_t^{ic}$ 
describes the mRNA transcripts that were initially present in the cell and still remain at 
time $t$, whereas $N_t^s$ describes the number of mRNAs 
transcribed since $t=0$.

Since $N_t^{ic}$ and $N_t^s$ are independent, it is easy 
to read off the first two moments directly:
\begin{align*} 
& \mean{N_t | \{\mu(\tau)\}_{\tau\in[0,t]},\{\lambda(\tau)\}_{\tau\in[0,t]},n_0} 
 = \mean{N_t^{ic}|n_0}+\mean{N_t^s} 
 = n_0\,e^{-\int_0^t \lambda(\tau) d\tau} + \chi(t);\\
& \R{Var}(N_t | \{\mu(\tau)\}_{\tau\in[0,t]},\{\lambda(\tau)\}_{\tau\in[0,t]},n_0) 
= \R{Var}(N_t^{ic}|n_0) + \R{Var}(N_t^s) 
= n_0e^{-\int_0^t \lambda(\tau) \, d\tau}  \left(1 - e^{-\int_0^t\lambda(\tau) \, d\tau} \right) + \chi(t). 
\end{align*}
From~\eqref{eq:NicNs}--\eqref{eq:Ns}, the full distribution is:
\begin{align}
 P(n,t | \{\mu(\tau)\}_{\tau\in[0,t]},\{\lambda(\tau)\}_{\tau\in[0,t]},n_0) 
&= \R{Pr}(N_t^{ic} + N_t^s=n | n_0) 
 =\sum_{k=0}^n \R{Pr}(N_t^{ic}=k|n_0)\,\R{Pr}(N_t^s=n-k) \nonumber \\
 &= \sum_{k=0}^n \binom{n_0}{k} \left(e^{-\int_0^t \lambda(\tau)d\tau}\right)^k
 \left(1-e^{-\int_0^t \lambda(\tau)d\tau}\right)^{n_0-k}\,
 \frac{\chi(t)^{n-k}}{(n-k)!}\,e^{-\chi(t)}.   \label{eq:MESolution}
\end{align}
This mathematical form is well-known when the rates are constant~\cite{Todorovic, Gardiner}, 
and a classical result in queueing theory~\cite{Eick:1993}.  We also remark that the solution 
with time-dependent rates~\eqref{eq:MESolution} is the one-gene
case of the main result in Jahnke and Huisinga~\cite{Jahnke:2007}.

If the initial state is itself described by a random variable $N_0$ with its
own probability distribution, we apply the law of total probability to 
obtain the solution in full generality as (see Appendix~\ref{app:N_ic}):
\begin{align}
\label{eq:PnPn0mix}
 P(n,t|\{\mu(\tau)\}_{\tau\in[0,t]},\{\lambda(\tau)\}_{\tau\in[0,t]}) 
 &=\sum_{n_0}P \left(n,t |\{\mu(\tau)\}_{\tau\in[0,t]},\{\lambda(\tau)\}_{\tau\in[0,t]},n_0 \right)
\,\R{Pr}(N_0=n_0) \nonumber \\
& = \sum_{k=0}^n \R{Pr}(N_t^s=n-k) \, \R{Pr}(N_t^{ic}=k),
\end{align}
where $N^s_t$ is distributed according to~\eqref{eq:NicNs}-\eqref{eq:Ns},
and $\R{Pr}(N_t^{ic}=k)$
is the mixture of the time-dependent binomial 
distribution~\eqref{eq:NicNs} and the distribution of the initial condition 
$N_0$. 

\paragraph*{\textbf{The initial transient `burn in' period.}}
For biologically realistic degradation rates $\{\lambda(t)\}_{t\ge 0}$,
the contribution from the initial condition ($N_t^{ic}$) 
decreases exponentially. Hence as time grows 
the transcripts present at $t=0$ degrade, and 
the population is expected to be composed 
of mRNAs transcribed after $t=0$. 

If the initial distribution of $N_0$ is not the stationary distribution of the ME
(or, more generally, not equal to the attracting distribution of the ME, as defined in 
Appendix~\ref{app:N_ic}),  
there is an initial time-dependence of $P(n,t)$ lasting over a time scale $T^{ic}$ 
(given by $\int_0^{T^{ic}} \lambda(\tau)\,d\tau \approx 1$), which
corresponds to a `burn-in' transient associated with the decay of the initial condition. 
We remark that the time-dependence described in~Refs.~\cite{Iyer-Biswas:2009, Shahrezaei:2008, Vandecan:2013, Smith:2015} corresponds only to this `burn-in' transient (see also~Fig.~\ref{fig:IB}).

On the other hand, when the initial distribution of $N_0$ is the stationary distribution
(or the attracting distribution) of the ME,
the component containing the initial condition ($N_t^{ic}$) and 
the long-term component ($N_t^{s}$) balance each other at every point
in time, maintaining stationarity (or the attracting distribution),
as shown analytically in Appendix~\ref{app:N_ic}.

\paragraph*{\textbf{The long-term behaviour of the solution.}}
In this work, we focus on the time dependence of $P(n,t)$ induced through 
non-stationarity of the parameters and/or correlated behaviour of the cells 
within the population. Hence for the remainder of the paper, we
neglect the transient terms.
Consequently, for perfectly synchronous cellular drives,
the solution of the ME~\eqref{eq:ME} is a Poisson random
variable with time-dependent rate equal to the effective upstream drive, 
$\chi(t)$:  $$[N_t | \chi(t)] \approx [N_t^s | \chi(t)] \sim \R{Poi}(\chi(t)),$$ 
with distribution
\begin{align}
\label{eq:PoiResult}
& P(n,t |\{\mu(\tau)\}_{\tau\in[0,t]},\{\lambda(\tau)\}_{\tau\in[0,t]})  = P(n,t | \chi(t)) = \frac{\chi(t)^n}{n!}e^{-\chi(t)},
\end{align}
which makes explicit the dependence on the \textit{history} of the sample paths
$\{\mu(\tau)\}_{\tau\in[0,t]}, \{\lambda(\tau)\}_{\tau\in[0,t]}$, which is encapsulated
in the \textit{value} of the effective drive $\chi(t)$ at time~$t$.

Indeed, from~\eqref{eq:x} it follows that 
the sample path $\{\chi(t)\}_{t\ge 0}$ satisfies a first order linear 
ordinary differential equation with time-varying coefficients:
\begin{equation}
\label{eq:xdiffeq}
 \frac{d \chi}{dt} + \lambda(t) \, \chi = \mu(t),
\end{equation}
which is the rate law for a chemical 
reaction with zeroth-order production with rate $\mu(t)$, and 
first-order degradation with rate $\lambda(t)$ per mRNA molecule. 
For biologically realistic (i.e., positive and 
finite) cellular drives, $\chi(t)$ is a continuous function.

\subsection{The general asynchronous case: cell-to-cell variability in the cellular drives}

Consider now the general case where different sample paths for the cellular
drives are possible, i.e., we allow explicitly for the 
transcription and degradation rates to vary from cell to cell.  The cell 
population will have some degree of asynchrony, hence
$M_t$ and $L_t$ have non-zero variance for at least some $t\ge0$.
The transcription and degradation rates 
are then described by stochastic processes $M$ and $L$:
\begin{align}
\label{reaction_diagram_random}
 \emptyset \xrightarrow{M_t} \text{mRNA} \xrightarrow{L_t} 
\emptyset,
\end{align}
and the collection of all differential equations of the 
form~\eqref{eq:xdiffeq} is represented formally by the random differential 
equation\footnote{We do not use the term \emph{stochastic} differential 
equation (SDE), because SDEs are usually associated with random white noise.}
\begin{equation}
\label{eq:Xdiffeq}
 \frac{dX_t}{dt} + L_t X_t = M_t. 
\end{equation} 
Equations of this form appear in many sciences, 
and a large body of classical results allows us to determine the 
probability density function of the upstream process, $X_t$~\cite{vanKampen, Soong, 
Pawula:1967}. Below, we use such results to 
obtain $f_{X_t}(x,t)$ for biologically relevant models.

Note that from Eq.~\eqref{eq:PoiResult} and the law of total probability, it follows 
that the probability mass  function for the random variable $N_t$ under cellular drives
described by the random processes $M$ and $L$ is given by 
the \emph{Poisson mixture} (or compound) distribution:
\begin{align}
P(n,t) = P_{X_t}(n,t) 
= \int \frac{x^n}{n!}\,e^{-x} \,f_{X_t}(x,t) \,dx,  \label{eq:PoiMixResult}
\end{align}
where the density $f_{X_t}(x,t)$ of the effective drive $X_t$ (to be determined)
can be understood as a \emph{mixing density}. The 
notation $P_{X_t}(n,t)$  recalls explicitly the dependence of the 
solution on the density of $X_t$, but we drop this reference
and use $P(n,t)$ below to simplify notation.
The problem of solving the full ME is thus
reduced to finding the mixing density $f_{X_t}(x,t)$.
Note that for synchronous drives, we have 
$f_{X_t}(x,t) = \delta(\chi(t)-x)$, where $\delta$ is the Dirac delta function, 
and~\eqref{eq:PoiMixResult} reduces to~\eqref{eq:PoiResult}.

Equation~\eqref{eq:PoiMixResult} also shows that there are two 
separate sources of variability in gene expression, contributing to the 
distribution of $N_t$. One source of variability is the Poisson nature of 
transcription and degradation, common to every model of the form 
considered here; the second source is the time-variation or uncertainty in the
cellular drives, encapsulated in the upstream process $X_t$
describing the `degree of synchrony' between cells and/or their variability over time.
In this sense, Eq.~\eqref{eq:PoiMixResult} connects with the concept of 
separable `intrinsic' and `extrinsic' components of gene expression noise
pioneered by Swain \emph{et al.}~\cite{Elowitz:2002, Swain:2002, 
Hilfinger:2011, Bowsher:2012}. Yet rather than considering moments, 
the full distribution $P(n,t)$ is separable into a model-dependent `upstream component' 
given by $f_{X_t}(x,t)$, and a downstream 
transcriptional `Poisson component' common to all models of this type.

\section{The effective upstream drive in gene transcription models }
\label{sec:fX}

The generic model of gene transcription and degradation with
time-dependent drives introduced above provides a unifying framework 
for several models previously considered in isolation.
In this section, we exemplify the tools to obtain 
the density of the effective drive $f_{X_t}(x,t)$ 
analytically or numerically through relevant examples.

\subsection{Gene transcription under upstream drives with static randomness}
\label{sec:static_drives}
In this first section, we consider models of gene transcription where the upstream drives are
deterministic, yet with random parameters representing cell variability.

\subsubsection{Random entrainment to upstream sinusoidal drives:
random phase offset in transcription or degradation rates}
\label{sec:sin}

Equation~\eqref{eq:Xdiffeq} can sometimes be solved directly to obtain 
$f_{X_t}(x,t)$ from a transformation of the random variables $M_t$ and $L_t$. 
We now show two such examples, where we explore the effect of entrainment
of gene transcription and degradation to an upstream periodic drive~\cite{Lueck:2014}.

First, consider a model of gene transcription of the form~\eqref{reaction_diagram_random} 
with transcription rate given by a sinusoidal function and where each cell has 
a random phase.  
This \emph{random entrainment} (RE) model is a simple representation of 
a cell population with transcription entrained to an upstream rhythmic signal, yet with a 
random phase offset for each cell:
\begin{equation}
\begin{aligned}
\label{eq:RE}
M_t & \defeq m\, \frac{1+\cos(\omega t + \Phi)}{2} \\ 
L_t & \defeq 1.
\end{aligned}
\end{equation}
Here $m$ and $\omega$ are given constants and 
$\Phi$ is a (static) random variable describing cell-to-cell variability (or uncertainty).
Solving Eq.~\eqref{eq:Xdiffeq} in this case, we obtain 
\begin{align*}
 X_t &= \frac{m \left(1+\omega^2 + \cos(\omega t +\Phi) + 
\omega \sin(\omega t + \Phi) \right)}{2(1+\omega^2)} \\
 &= B + A \, \sin(\omega t + \Phi^*),
\end{align*}
where $A = m/2\sqrt{1+\omega^2}$, $B = m/2$ and 
$\Phi^* = \Phi+\arctan(1/\omega)$.

\begin{figure*}[htb!]
\centerline{\includegraphics[width=\textwidth]{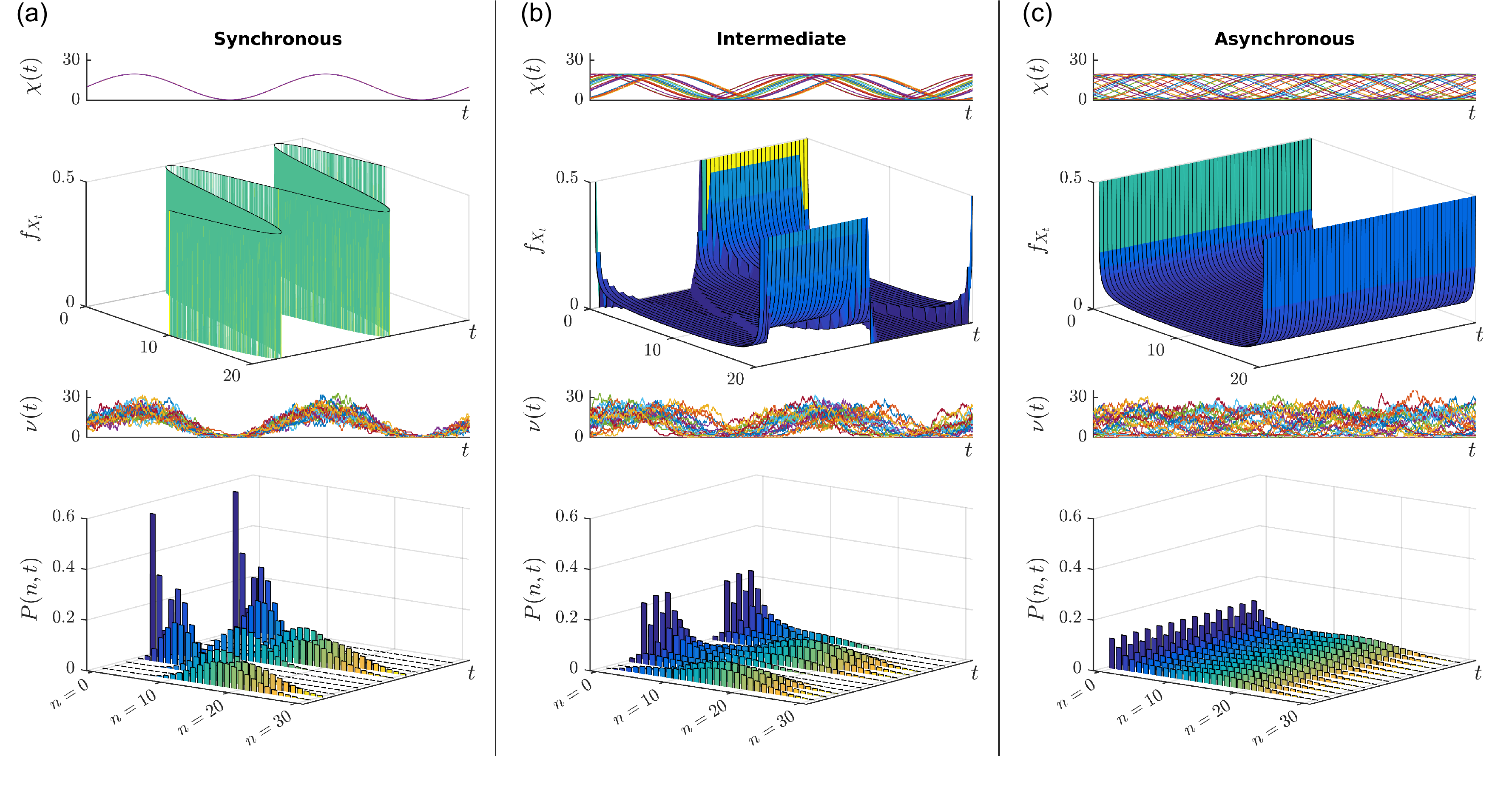}}  
\caption{\label{fig:sin}
  Gene transcription under the random entrainment model~\eqref{eq:RE}
  with constant degradation rate and transcription rates entrained to an 
  upstream sinusoidal signal with $\omega =1$.
  Each cell has a random phase offset $\phi$ drawn from a distribution. 
(a) The synchronous population corresponds to identical
phases across the population. In this case, the transcription 
reflects the time variability of the upstream drive mixed with   
the stochasticity due to the downstream Poisson process of transcription.
When the random phases $\phi$ 
are uniformly distributed on an interval of range (b) $\pi$ and (c) $2 \pi$, 
the population becomes increasingly asynchronous. 
For all three cases, we show (top to bottom): sample paths of 
the effective drive, $X$; its density $f_{X_t}(x,t)$ given by Eq.~\eqref{eq:fX_sine};
sample paths of the number of mRNAs, $N$; and the full solution of the ME $P(n,t)$.
}
\end{figure*}

Suppose $\Phi^*$ is uniformly distributed on $[-r,r]$, $r \le \pi$.
Inverting the sine with $\Phi^*$ restricted to 
$[-r,r]$, we obtain 
\begin{align}
\label{eq:fX_sine}
 f_{X_t}(x,t) &= \frac{k(t)}{2r\sqrt{A^2-(x-B)^2}},
\end{align}
where $k(t)\in\{0,1,2\}$
is the number of solutions of 
$\sin \theta=(x-B)/A$ for $\theta \in (\omega t-r, \omega t + r)$. 
As the phase distribution of the drives becomes narrower,
the upstream variability disappears: $r \to 0 \implies f_{X_t}(x,t) \to 
\delta((B+A\sin \omega t) - x)$. In this limit, all cells follow the entraining drive exactly, 
and $P(n,t)$ becomes a Poisson distribution at all times.

Figure~\ref{fig:sin} depicts $f_{X_t}$ for $r=0$ (no cell-to-cell phase 
variation, (a)) and for $r=\pi/2$, and $r=\pi$ 
(increasingly wider uniform distribution of phases, (b) and (c)). 
The full distribution $P(n,t)$ is obtained 
using~\eqref{eq:fX_sine}~and~\eqref{eq:PoiMixResult}.

Second, let us consider the same model of entrainment to
an upstream sinusoidal signal with a random offset, but this time via the 
degradation rate:
\begin{equation}
\begin{aligned}
\label{eq:REdeg}
M_t & \defeq m, \\ 
L_t & \defeq b+ a\cos(\omega t + \Phi),
\end{aligned}
\end{equation}
with $m$, $a$, $b$, and $\omega$ given constants, and 
$\Phi$ a (static) random variable.

Eq.~\eqref{eq:Xdiffeq} can be solved approximately~\cite{Lueck:2014} to give
\begin{equation*}
 X_t = B + A \, \sin(\omega t + \Phi^*),
\end{equation*}
where $A = 2ma\sqrt{1+(\omega/b)^2}/[2(b^2+\omega^2)-a^2]$, 
$B =Ab\sqrt{1+(\omega/b)^2}/a$, and 
$\Phi^* = \Phi+ \pi + \arctan(b/\omega)$.
As before, if $\Phi^*$ is uniform on  $[-r,r]$, $r \le \pi$,
the density of the effective drive takes the same form~\eqref{eq:fX_sine} as
above.

\subsubsection{Upstream Kuramoto promoters with varying degree of 
synchronisation}
\label{sec:KPM}
As an illustrative computational example,
we study a population of cells whose promoter strengths display a 
degree of synchronisation across the population.
To model this upstream synchronisation, consider the \emph{Kuramoto promoter 
model},
where the promoter strength of each cell $i$ depends on an oscillatory phase 
$\theta_i(t)$, and 
cells are coupled via a Kuramoto model~\cite{Kuramoto:1975, 
Strogatz:2000, jadbabaie2004stability}. 
We then have a model of the form~\eqref{reaction_diagram_random} with:
\begin{equation}
 \begin{aligned}
\label{eq:kuramoto}
M_t & \defeq m \,\frac{b + \cos \left(\Theta(t; \Omega)\right)}{2} \\
L_t & \defeq  1.
\end{aligned}
\end{equation}
Here $m, b$ are constants and $\{\theta_i(t)\}_{i=1}^C$ 
are the phase variables for the $C$ cells
governed by	 the globally coupled Kuramoto model:
\begin{equation}
\label{eq:kuramoto2}
    \frac{d \theta_i}{d t} = \omega_i + \frac{K}{C} \sum_{j=1}^C\sin(\theta_j - \theta_i),
\end{equation}
where $K$ is the coupling parameter and 
the intrinsic frequency of each cell, $\omega_i$, is drawn from the random distribution 
$\Omega \sim \mathcal{N}(0,0.05^2)$. 
The Kuramoto model allows us to tune the degree of synchrony
through the coupling $K$: for small $K$, the cells 
do not display synchrony since they all have a slightly different
intrinsic frequency; as $K$ is increased, the population
becomes more synchronised.

This model is a simple representation of nonlinear synchronisation processes 
in cell populations with intrinsic 
heterogeneity~\cite{Garcia-Ojalvo:2004, Taylor:2009, Murray:2011, Uriu:2012}.
In Figure~\ref{fig:kuramoto}(a), 
we show how the sample paths change
as the degree of synchrony increases, and we exemplify the use of~\eqref{eq:PoiMixResult}
for the numerical solution of the gene expression of this model.

\subsection{Asynchronous transcription under stochastic multistate promoters}
\label{sec:multistate}

In the previous section, we obtained $f_{X_t}(x,t)$ by capitalising on the 
precise knowledge 
of the sample paths of $M$ and $L$ to solve~\eqref{eq:Xdiffeq} explicitly.
In other cases, we can obtain $f_{X_t}(x,t)$ by following the usual procedure of 
writing down an evolution equation for the probability density of 
an \emph{expanded} state that is Markovian, and then marginalising. 
More specifically, let the vector process $\mathbf{Y}$ prescribe the upstream drives, 
so that $M = M(\mathbf{Y},t)$ and $L = L(\mathbf{Y},t)$,
and consider the expanded state $\mathbf{X}_t\defeq (X_t, \mathbf{Y}_t )$.
Note that since $\mathbf{Y}$ is upstream, it prescribes $X$ (and not vice versa).
We can then write the evolution equation for the joint probability density 
$f_{\mathbf{X_t}}(x,\mathbf{y},t)$: 
 \begin{align}
\label{eq:evolf}
 \frac{\partial}{\partial t} f_{\mathbf{X_t}}(x,\mathbf{y},t) 
 = &-\frac{\partial}{\partial x} \left[ \left(\mu(\mathbf{y},t)-\lambda(\mathbf{y},t) x \right) 
 \, f_{\mathbf{X_t}}(x,\mathbf{y},t) \right]
 + \mathcal{L}_{\mathbf{Y_t}} \left[ f_{\mathbf{X_t}}(x,\mathbf{y},t) \right],
\end{align}
which follows from conservation of probability.
In Equation~\eqref{eq:evolf}, 
the differential operator for $X$, which follows from~\eqref{eq:Xdiffeq}, 
is the first jump moment~\cite{Pawula:1970}
conditional upon $\mathbf{Y_t}=\mathbf{y}$ (and hence upon 
$M_t = \mu(\mathbf{y},t)$ and $L_t = \lambda(\mathbf{y},t)$);
the second term $\mathcal{L}_{\mathbf{Y_t}}[.]$ is the infinitesimal generator
of the upstream processes.
In particular, for continuous stochastic processes
$\mathcal{L}_{\mathbf{Y_t}}[.]$ is of Fokker-Planck type,
and for Markov chains $\mathcal{L}_{\mathbf{Y_t}}[.]$ is a transition rate 
matrix.
The desired density $f_{X_t}(x,t)$ can then be obtained
via marginalisation.

\begin{figure}[htb!]
 \centerline{\includegraphics[width=.5\textwidth]{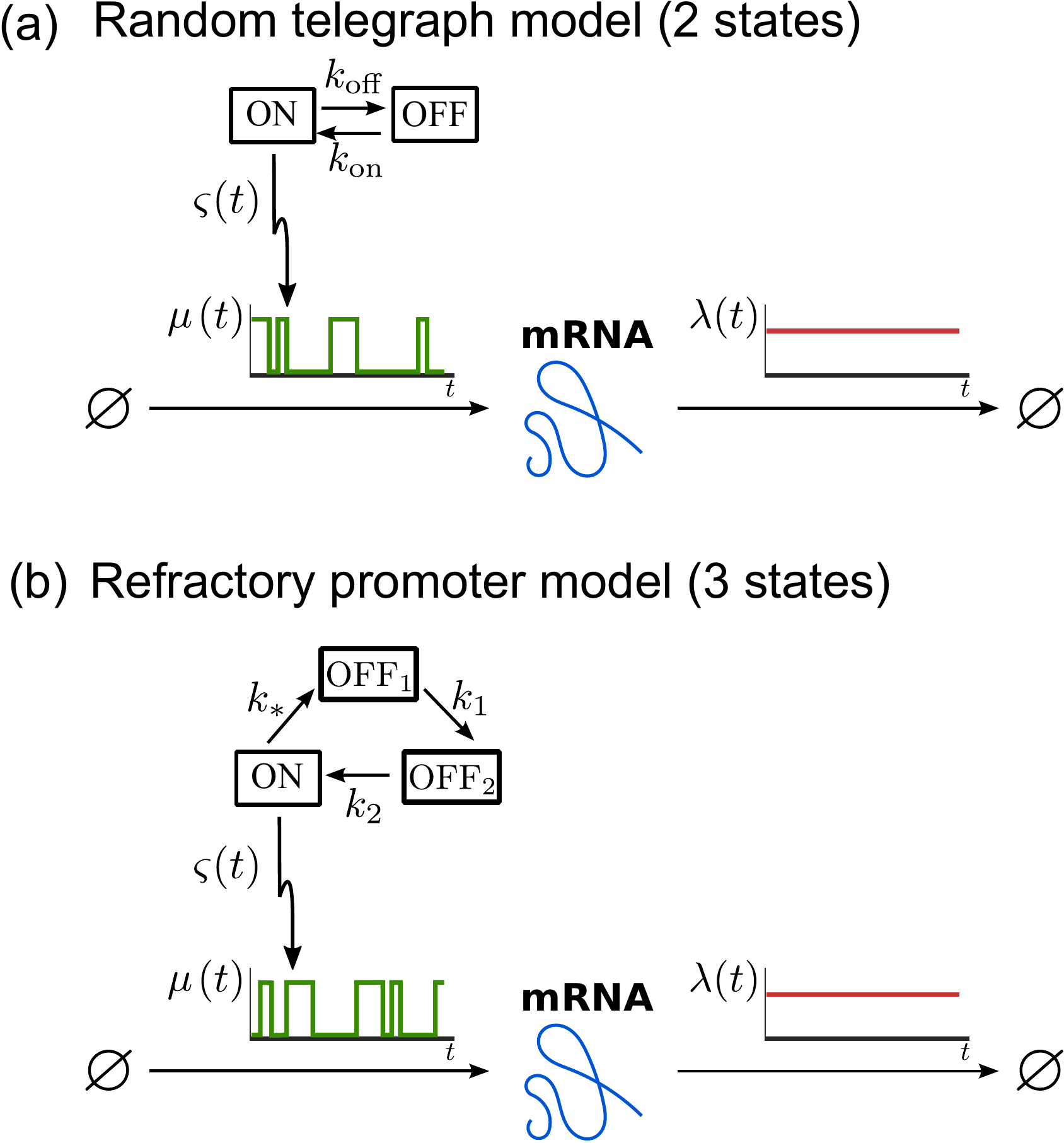}}
 \caption{Asynchronous stochastic promoter switching models correspond to 
upstream stochastic processes. The promoter cycles between the discrete 
 states, transitioning stochastically with 
rates as indicated: (a) the standard 2-state random 
telegraph (RT) model; (b) the 3-state refractory promoter model.}
\label{fig:cycles}
\end{figure}

Equation~\eqref{eq:evolf} can be employed to analyse the widely used class of
transcription models with asynchronous, random promoter switching between discrete 
states, where each state has different transcription and degradation 
rates representing different levels of promoter 
activity due to, e.g., transcription factor binding or chromatin 
remodelling~\cite{Sanchez:2011}.
A classic example is the \emph{random telegraph} (RT) model, 
first used by Ko in 1991~\cite{Ko:1991} to explain cell-to-cell 
heterogeneity and bursty transcription (Fig.~\ref{fig:cycles}a).

In our framework, such random promoter switching can be understood 
as an upstream \emph{stochastic} process driving transcription as follows.
Let us assume that the promoter can attain $D$ states $s$,
and each state has constant transcription rate 
$m_s$ and constant degradation rate $\ell_s$.
The state of the promoter is described by a random process
$S=\{S_t \in \{1,2,\dots,D\}: \,t\ge 0\}$, with sample paths 
denoted by $\{\varsigma(t)\}_{t\ge 0}$, and its evolution 
is governed by the $D$-state Markov chain 
with transition rate $k_{sr}$ from state $r$ to state $s$. 
The state of the promoter $S_t=s$ 
prescribes that $M_t=m_s$ and $L_t = \ell_s$. 
Hence, the sample paths of $M$ and $L$ are a succession of step functions
with heights $m_s$ and $\ell_s$, respectively,
occurring at exponentially distributed random times. 

As described above, we expand the state space of the cellular drives 
to include the promoter state $\mathbf{X}_t = \{X_t, S_t\}$.
The evolution equation~\eqref{eq:evolf} is then given by $D$ coupled equations:
\begin{multline}
\label{eq:multKM}
 \frac{\partial}{\partial t}f_{X_t,S_t}(x,s,t) 
 = -\frac{\partial}{\partial x}\left[(\mu_s-\lambda_s x) 
f_{X_t,S_t}(x,s,t)\right] 
 + \sum_{j=1}^D  k_{sj} f_{X_t,S_t}(x,j,t) - \sum_{j=1}^D 
 k_{js} f_{X_t,S_t}(x,s,t) ,  \\ 
\quad \quad  s=1,2, \ldots, D,
\end{multline}
which can be thought of as a set of multistate Fokker-Planck-Kolmogorov 
equations~\cite{Pawula:1970}.
Marginalisation then leads to the density of the effective drive:
\begin{align}
\label{eq:sum_of_states}
f_{X_t}(x,t) = \sum_{s=1}^D  f_{X_t,S_t}(x,s,t),
\end{align} 
and the full ME solution is obtained from~\eqref{eq:sum_of_states} 
and~\eqref{eq:PoiMixResult}.

We illustrate this approach more explicitly with two examples (Fig.~\ref{fig:cycles}): 
a re-derivation of the known solution of the standard RT model; 
and the solution of the 3-state cyclic model with a refractory state. 
Results for other promoter architectures are discussed in~\cite{thesis}.

\subsubsection{The Random Telegraph model (2 states)}

Although the RT model has been solved by several 
methods~\cite{Peccoud:1995, Raj:2006, Iyer-Biswas:2009}, 
we briefly rederive its solution within the above framework 
to clarify its generalisation to other promoter architectures.

Consider the standard RT model (Fig.~\ref{fig:cycles}a), with promoter 
switching stochastically between the active state $s_{\R{on}}=1$, with 
constant transcription rate $m_1=m$, and the inactive state $s_{\R{off}}=0$, where no 
transcription takes place, $m_0 =0$. 
The transition rates between the two states are $k_{10}=k_{\R{on}}$ and 
$k_{01}=k_{\R{off}}$. Without loss of generality, we 
assume $\ell_1 = \ell_0 = \lambda(t) \equiv 1$.  The transcription model 
is of the 
form~\eqref{reaction_diagram_random} with:
\begin{equation}
 \begin{aligned}
\label{eq:RT}
M_t & \defeq m \, S_t \\
L_t & \defeq  1,
\end{aligned}
\end{equation}
where $S=\{S_t \in \{0,1\}: \,t\ge 0\}$  with waiting times drawn from
exponential distributions: $\tau_{\R{off}} \sim \exp(1/k_{\R{on}})$ and 
$\tau_{\R{on}} \sim \exp(1/k_{\R{off}})$.

Let $Z_t=X_t/m$, and denote
$f_{\R{on}}(z,t) \defeq f_{Z_t,S_t}(z,s_{\R{on}},t)$ and 
$f_{\R{off}}(z,t) \defeq f_{Z_t,S_t}(z,s_{\R{off}},t)$, with $z \in (0,1)$. 
Then the multistate Fokker-Planck-Kolmogorov equations~\eqref{eq:multKM} are:
\begin{align*}
 \frac{\partial f_{\R{on}}}{\partial t} & = 
 -\frac{\partial}{\partial z}\left[(1-z)f_{\R{on}}\right] 
-k_{\R{off}}f_{\R{on}}+k_{\R{on}}f_{\R{off}} \\
 \frac{\partial f_{\R{off}}}{\partial t} &= 
 -\frac{\partial}{\partial z}\left[-zf_{\R{off}}\right] + 
k_{\R{off}}f_{\R{on}}-k_{\R{on}}f_{\R{off}}, \\
f_{Z_t} & =f_{\R{on}}+f_{\R{off}}
\end{align*}
with integral conditions $\int_0^1 f_{\R{on}}(z,t)\D z = P(S_t=s_{\R{on}})$ 
and $\int_0^1 f_{\R{off}}(z,t)\D z = P(S_t=s_{\R{off}})$.

At stationarity, it then follows~\cite{Smiley:2010} that 
\begin{equation}
\label{eq:fZRT}
f_{Z_t}(z)=\frac{z^{k_{\R{on}}-1}(1-z)^{k_{\R{off}}-1}}{\R{B}
\left(k_{\R{on}},k_{\R{off}}\right)},
\end{equation}
where $\R{B}(a,b) = \Gamma(a)\Gamma(b)/\Gamma(a+b)$ 
is the Beta function. In other words, at stationarity, 
the normalised effective drive is described by a Beta distribution: 
\begin{align*}
Z_t\sim\R{Beta}\left(k_{\R{on}},k_{\R{off}}\right), \quad \forall t.
\end{align*}
Using~\eqref{eq:fZRT}~and~\eqref{eq:PoiMixResult}, we
obtain that the full stationary solution is the Poisson-Beta mixture:
\begin{align}
P(n) &= \int_0^1 \frac{(m z)^n}{n!}\, e^{-m z}\, 
\frac{z^{k_{\R{on}}-1}(1-z)^{k_{\R{off}}-1}}{\R{B}
\left(k_{\R{on}},k_{\R{off}}\right)}\D z \nonumber \\
\label{eq:RTsoln}
 &= \frac{\Gamma\left(k_{\R{on}}+n\right)}{\Gamma
\left(k_{\R{on}}\right)}
\frac{\Gamma\left(k_{\R{on}}+k_{\R{off}}\right)}{\Gamma\left(k_{\R{on}}+k_{\R{off}}+n\right)} \,\,
\frac{m^n}{\Gamma\left(n+1\right)}  \,\, \tensor*[_1]{F}{_1}\left(k_{\R{on}}+n,k_{\R{on}}+k_{\R{off}}
+n;-m\right),
\end{align}
where $\tensor*[_1]{F}{_1}\left(a,b;z\right)$ is the confluent 
hypergeometric function~\cite{Abramowitz}.

\subsubsection{The Refractory Promoter model (3 states) }
\label{sec:3state}

In the standard RT model, the waiting times in each state are exponentially distributed.  
In recent years, time course data have shown that the $\tau_\text{off}$ do
not conform to an exponential distribution, leading some authors to 
incorporate a second inactive (refractory) state, which needs to be cycled 
through before returning to the active state~\cite{Suter:2011, Zoller:2015}. 
The net `OFF' time is then the sum of two exponentially distributed waiting times.

In this \textit{refractory promoter model} (Fig.~\ref{fig:cycles}b), 
the promoter switches through the states $s_*$, $s_1$, and 
$s_2$ with rates $k_*$, $k_1$ and $k_2$.  Transcription takes place at constant rate $m$ 
only when the promoter is in the active state $s_*$ and, without loss of generality, we assume
a constant degradation rate  $\lambda(t) \equiv 1$ for all states.
This model is of the same form as~\eqref{eq:RT}, and is solved similarly.

Making the change of variables $Z_t= \lambda X_t/m=X_t/m$ 
and, using the notation $f_{i}(z,t) \defeq  f_{Z_t,S_t}(z,t, s_i)$,
the multistate Fokker-Planck-Kolmogorov equations are
\begin{align*}
 \frac{\partial f_*}{\partial t} & = 
 -\frac{\partial}{\partial z}\left[(1-z)f_*\right] -k_*f_*+k_2f_2 \\
 \frac{\partial f_1}{\partial t} & = 
 -\frac{\partial}{\partial z}\left[-zf_1\right] + k_*f_*-k_1f_1 \\
 \frac{\partial f_2}{\partial t} & = 
 -\frac{\partial}{\partial z}\left[-zf_2\right] + k_1f_1-k_2f_2,  \\
 f_{Z_t}  & = f_* + f_1 +  f_2. 
\end{align*}
with three integral conditions $\int_0^1 f_i(z) \D z = P(S=s_i)$.

At stationarity, we find
\begin{align*}
f_{Z_t}(z) &= 
C_1\;z^{k_1-1}\;\tensor*[_2]{F}{_1}\left[a^{(1)}_+,a^{(1)}_-;1+k_1-k_2;  z\right] 
+ C_2\;z^{k_2-1}\; \tensor*[_2]{F}{_1}\left[a^{(2)}_+,a^{(2)}_-;1-k_1+k_2; z\right], \\
& a^{(1)}_{\pm} \defeq \frac{1}{2}\left(2+k_1-k_2-k_*\pm d\right), \quad
  a^{(2)}_{\pm} \defeq \frac{1}{2}\left(2-k_1+k_2-k_*\pm d \right),  \nonumber 
\end{align*}
where $\{k_*, \, (k_{1}-k_{2}), \, d \defeq \sqrt{(k_*-k_1-k_2)^2-4k_1k_2} \} \notin \mathbb{Z}$ 
and $\tensor*[_2]{F}{_1}\left(a,b;c;z\right)$ is the 
Gauss hypergeometric function~\cite{Abramowitz}.
The full stationary solution $P(n)$ is then obtained from~\eqref{eq:PoiMixResult}.

For a detailed derivation (including expressions for the integration 
constants $C_1$ and $C_2$), see Appendix~\ref{app:3state}.

\subsection{Asynchronous multistate models with upstream promoter modulation}
\label{sec:multimodulation}

Finally, we consider a model of gene transcription 
that incorporates features of models described in
Sections~\ref{sec:static_drives}~and~\ref{sec:multistate}.
Such a situation is of biological interest and appears when individual cells 
exhibit correlated dynamics in response to upstream factors
(e.g., changing environmental conditions, drives or stimulations),
but still maintain asynchrony in internal processes, such as transcription factor 
binding~\cite{Zechner:2014, Hasenauer:2014}.

\begin{figure}[htb!]
 \centerline{\includegraphics[width=.7\textwidth]{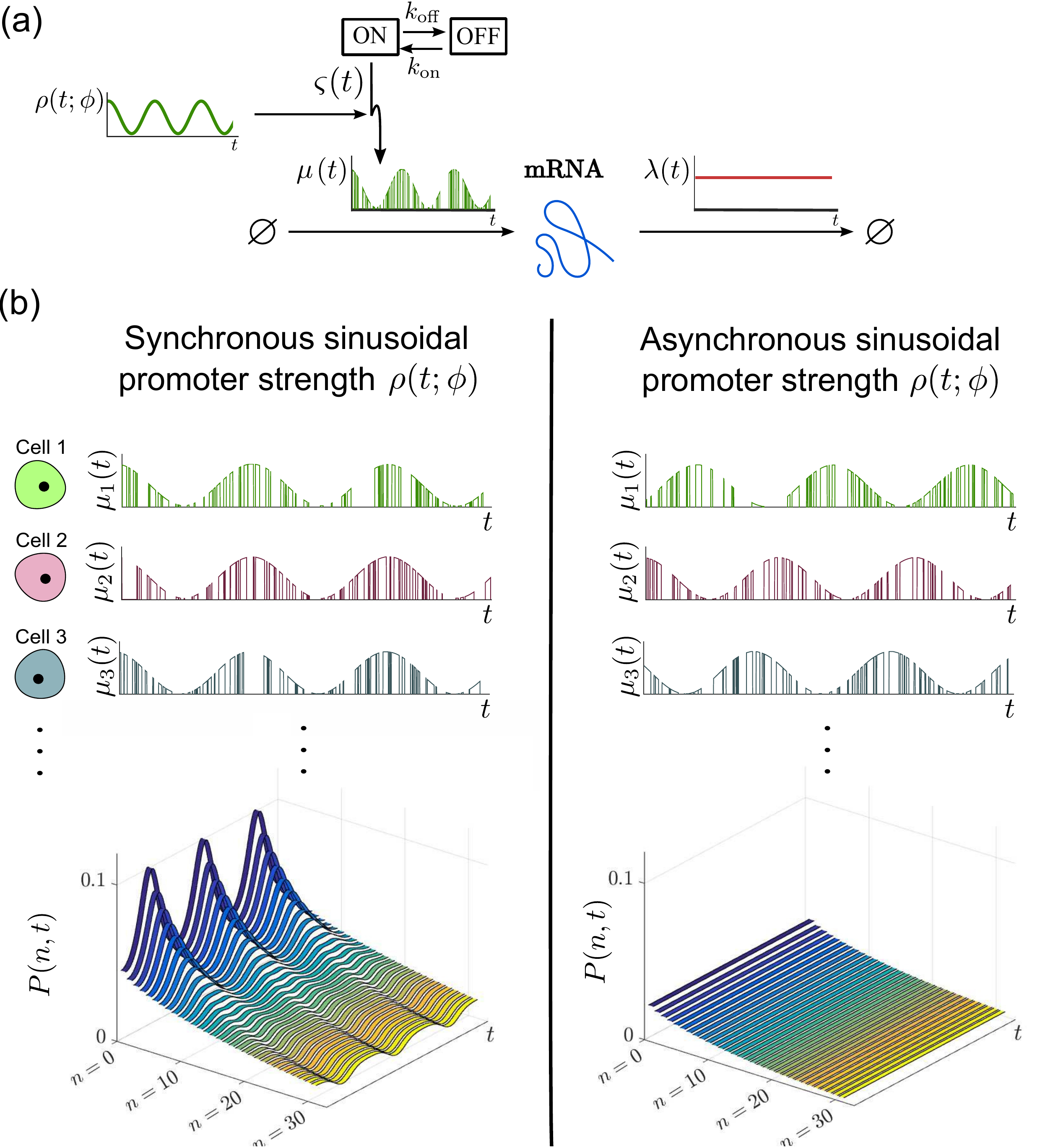}}
 \caption{(a) Modulated Random Telegraph (MRT) model: each cell switches
 asynchronously between `ON' and `OFF' states, but the magnitude of the 
 `ON' transcription rate is modulated by the function $\rho(t; \phi)$,
 a sinusoid representing an upstream periodic
 process. The phase $\phi$ represents the cell-to-cell variability
 and leads to the varying degree of synchrony across the population.
(b) Sample paths $\mu_i(t)$ and solution of the probability distribution $P(n,t)$ 
of the MRT model for synchronous (left) and asynchronous (right) modulation.
In the asynchronous case, the upstream drive has a random phase
across the cells with distribution $\Phi \sim \mathcal{N}(0, 10)$. 
}
\label{fig:RTsin}
\end{figure}

To illustrate this concept, we consider the \textit{modulated random telegraph} (MRT) model, 
a combination of the RE model~\eqref{eq:RE} and the RT model~\eqref{eq:RT}, 
i.e., the promoter strength is modulated by an upstream sinusoidal drive with random 
phase $\Phi$, as in the RE model, yet the promoter switches 
stochastically between active/inactive states with rates $k_{\R{on}}$ 
and $k_{\R{off}}$, as in the RT model. 
In this model, the transcription rate is correlated between cells 
through the entrainment to the upstream sinusoidal drive as a 
simple model for, e.g., circadian gene expression:
\begin{align*}
M_t & \defeq  R_t(\Phi) \, S_t  =  m \frac{1+\cos(\omega t+\Phi)}{2} S_t \\
L_t & \defeq  1,
\end{align*}
where $m, \omega>0$ are 
constants; $\Phi$ is the random phase across the cell population; and 
$S=\{S_t \in \{0,1\}: \,t\ge 0\}$, with exponential waiting times, 
describes the stochastic promoter switching (Fig.~\ref{fig:RTsin}a).

The solution of this model follows from the RT probability density~\eqref{eq:fZRT} 
conditioned on the random phase $\Phi$,
which prescribes the sample path 
$\{\rho(t; \phi)\}_{t\ge 0}$ of the promoter strength $R$.
The resulting scaled Beta distribution
\begin{align*}
 f_{X_t|\Phi}(x,t|\phi) 
 = \frac{x^{k_{\R{on}}-1}(\rho(t; \phi)-x)^{k_{\R{off}}-1}}{\R{B}(k_{\R{on}},
k_{\R{off}}) \, \rho(t; \phi)^{k_{\R{on}}+k_{\R{off}}-1}}
\end{align*}
is then marginalised over the phase $\Phi$ to obtain the density $f_{X_t}(x,t)$ 
of the effective drive.
For instance, if the phases are normally distributed 
$\Phi \sim\mathcal{N}(0,\sigma^2)$, we have (see Fig.~\ref{fig:RTsin}b): 
\begin{align*}
 f_{X_t}(x,t) = \int f_{X_t|\Phi}(x,t|\phi) \, f_{\Phi}(\phi) \D\phi 
 = \int_{-\infty}^{\infty} 
\frac{x^{k_{\R{on}}-1}}{\R{B}(k_{\R{on}},k_{\R{off}})} \,
 \frac{\left(\frac{m}{2}[1+ \cos(wt+\phi)]-x\right)^{k_{\R{off}}-1}}
 {\left(\frac{m}{2}[1+ \cos(wt+\phi)]\right)^{k_{\R{on}}+k_{\R{off}}-1}} \,\,
\frac{e^{-\phi^2/2\sigma}}{\sigma\sqrt{2\pi}}\, \D\phi.
\end{align*}
As $\sigma\to \infty$, the population becomes asynchronous in the 
promoter strength, as well as in the state transitions, and time dependence wanes 
(Fig.~\ref{fig:RTsin}b).

\section{Ensemble noise characteristics in time-varying populations}
\label{sec:time_population}

In the previous sections, we were concerned with the full
time-dependent probability distribution $P(n,t)$ for the mRNA 
copy number $N$. However, in many circumstances such detailed 
information is not required, and simpler characterisations based 
on ensemble averages (e.g., Fano factor, coefficient of variation) are of interest.
Simple corollaries from the Poisson mixture expression~\eqref{eq:PoiMixResult}
allow us to derive expressions for the ensemble moments and other noise 
characteristics, as shown below. We remark that, in this section, 
all the expectations are taken over the distribution describing the cell population.

\subsection{Time-dependent ensemble moments over the distribution of cells}

To  quantify noise characteristics of gene expression in a population, 
the ensemble moments $\mean{N_t^k}, \  k \in\mathbb{N}, \ t\ge 0$ 
are often determined via the probability generating 
function~\cite{Peccoud:1995, Paszek:2007, Zhang:2012} or by integrating the 
master equation~\cite{Sanchez:2008, Sanchez:2011, Sanchez:2013}.
However, stationarity is usually assumed and
the moments derived are not suitable for time-varying systems. 
Here we use corollaries of the Poisson mixture 
expression~\eqref{eq:PoiMixResult} to derive expressions for the ensemble 
moments for time-varying systems under upstream drives.

From~\eqref{eq:PoiResult} we have
$[N_t|X_t=x] \sim \R{Poi}(x)$; hence
\begin{align*}
&\mean{N_t^k|X_t=x} = \sum_{r=1}^k x^r S(k,r),
\quad \text{where}  \quad
 S(k,r) = \sum_{j=0}^r \binom{r}{j}\frac{(-1)^{r-j}j^k}{r!}
\end{align*}
are the Stirling numbers of the second kind~\cite{Riordan:1937}.
The law of total probability then gives
\begin{equation}
 \mean{N_t^k} = \sum_{r=1}^k S(k,r)\,\mean{X_t^r},
 \label{eq:moments}
\end{equation}
or, equivalently,
\begin{equation}
\begin{small}
 \begin{pmatrix}
  \mean{N_t} \\
  \mean{N_t^2} \\
  \vdots \\
  \mean{N_t^k}
 \end{pmatrix}
=\begin{pmatrix}
  S(1,1) &  0     & \dots & 0 \\
  S(2,1) & S(2,2) & \dots & 0 \\
  \vdots & \vdots & \ddots & \vdots \\
  S(k,1) & S(k,2) & \dots & S(k,k)
 \end{pmatrix}
  \begin{pmatrix}
  \mean{X_t} \\
  \mean{X_t^2} \\
  \vdots \\
  \mean{X_t^k}
 \end{pmatrix}. \nonumber
\end{small}
\end{equation}

Therefore the ensemble moments of the mRNA copy number $\mean{N_t^k}$ can be
obtained in terms of the moments of the effective drive $\mean{X_t^k}$, and vice versa. 
For instance, it follows easily that
$\mean{X_t}=\mean{N_t}$; 
$\R{Var}(X_t)=\R{Var}(N_t)-\mean{N_t}$; and the skewness $\gamma_1(X_t) = 
(\mean{N_t^3}-3\R{Var}(N_t)-3\mean{N_t}\R{Var}(N_t)-\mean{N_t}^3)/(\R{Var}(N_t)-
\mean{N_t})^{3/2}$. 

\paragraph*{\textbf{Decomposing the sources of noise:}}
From Eq.~\eqref{eq:moments}, it follows that the variability of the mRNA count $N_t$ 
can be rewritten as:
\begin{align}
 \eta_{N}^2(t) \defeq \frac{\R{Var}(N_t)}{\left(\mean{N_t}\right)^2}
 &= \frac{1}{\mean{N_t}} + \frac{\R{Var}(X_t)}{\left(\mean{X_t}\right)^2} 
 \label{eq:noisedecomp}
 = \eta^2_{\R{Poi}}(t) + \eta^2_{\R{up}}(t),
\end{align}
i.e., it can be decomposed into a Poissonian (downstream) component $\eta^2_{\R{Poi}}(t)$ and 
an upstream component $\eta^2_{\R{up}}(t)$ linked to the variable $X_t$. 
Note, however, that our expressions~\eqref{eq:moments} provide decompositions for all moments, and not 
only the mean and variance.

Expression~\eqref{eq:noisedecomp} can be mapped onto the common decomposition 
into `intrinsic' and  `extrinsic' components~\cite{Swain:2002, Hilfinger:2011}, if we note that, 
in our model, the
`intrinsic' components are the downstream processes of transcription and degradation, whose 
rates are affected by the `extrinsic' variability due to upstream factors.  Such upstream factors
can be biologically diverse, and can be both intra- and extra-cellular.
Therefore throughout this paper we refer to `upstream/downstream' processes instead
of `intrinsic/extrinsic' noise, to emphasise that upstream processes can 
reflect variability that is internal to the cell as well as cell-to-cell variability.
For example, the asynchronous stochastic promoter switching described in
Sec.~\ref{sec:multistate} is an upstream process here, which in the literature might have been classed as `intrinsic'
(although in fact, asynchronicity implies an assumption about cell-to-cell variability).
On the other hand, the modulated promoter switching in Sec.~\ref{sec:multimodulation} includes 
both `intrinsic' and `extrinsic' sources of variability, as usually classed in the literature.  
In our framework, such processes are treated consistently as `upstream' sources of variability.

\paragraph*{\textbf{Analysis of time-dependent moments:} }

The relationship~\eqref{eq:moments}
between downstream and upstream moments together with the dynamical 
equation~\eqref{eq:Xdiffeq} enables us to solve for the time-dependence of the
moments of mRNA counts in terms of the moments of the drive: 
\begin{align}
\label{eq:ENXM}
 &\mean{N_t} = \mean{X_t} = \int_0^t e^{-\lambda(t-\tau)}\mean{M_{\tau}}\,d\tau; \\
 &\mean{N_t^2} = \mean{X_t} + \mean{X_t^2} 
 =\mean{N_t} + \int_0^t \int_0^t  e^{-\lambda(2t-\tau-\sigma)} 
\mean{M_{\tau}M_{\sigma}}\,d\tau\, d\sigma,
\label{eq:ENXM2}
\end{align}
where for simplicity we have assumed constant degradation rate $\lambda$.
(For the most general case with degradation rate $L_t$, see for example Ref.~\cite{Soong}.) 
Therefore, the observed moments $\mean{N_t^k}$ from the data can be used to
infer the time-dependent moments of the (usually unobserved) upstream drives.

\begin{figure}[htb!]
 \centerline{\includegraphics[width=.75\textwidth]{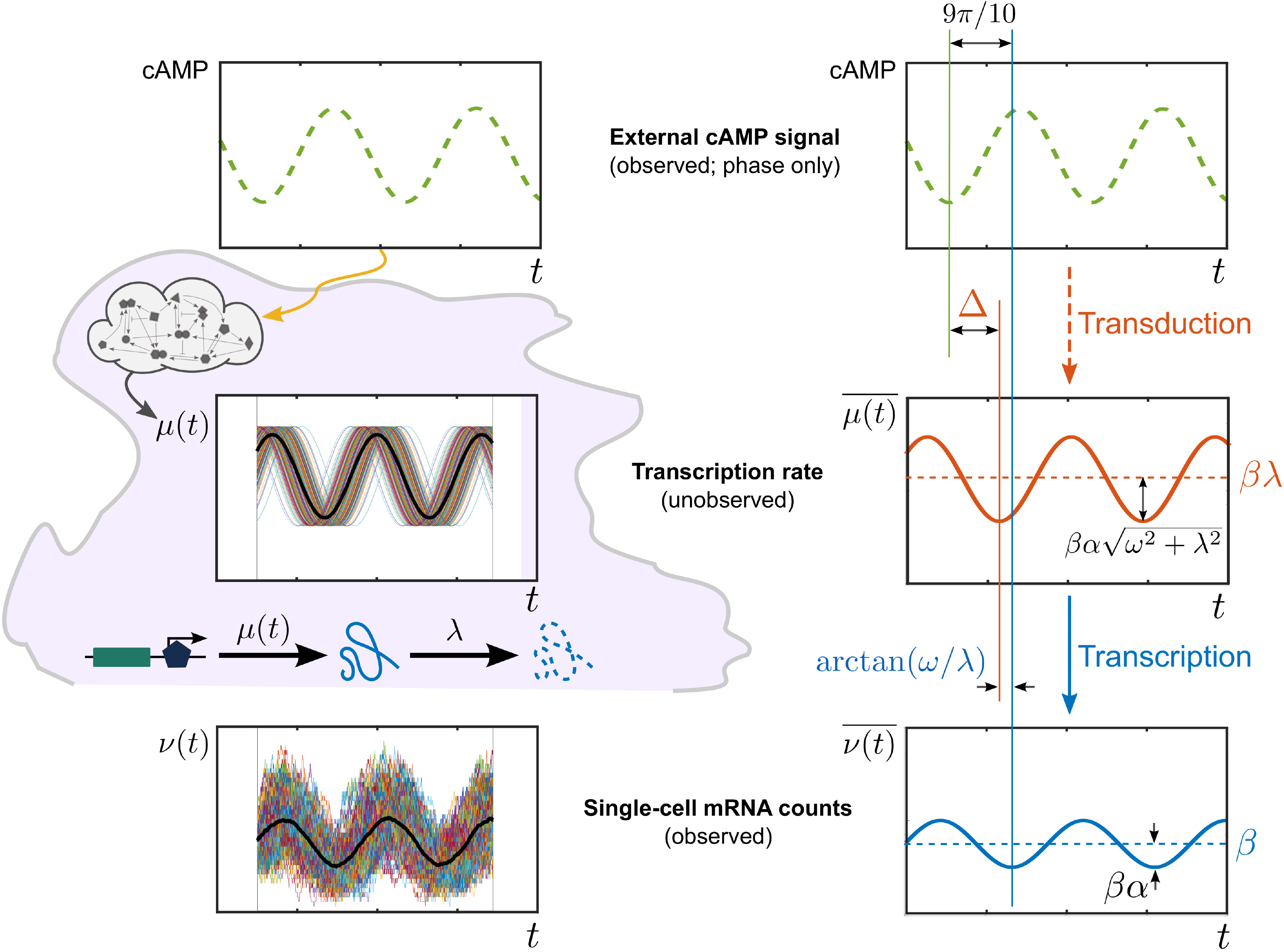}}
 \caption{Analysis of single-cell temporal transcription of a gene
 in response to an upstream oscillatory cAMP signal, motivated by recent experiments~\cite{Corrigan:2014}. 
 Individual single-cell time courses $\nu(t)$ of mRNA counts are highly variable with no clear
 entrainment to the driving signal, whereas the time-dependent ensemble average $\overline{\nu(t)}$ 
 oscillates with the same frequency as the external drive. This is consistent with Eqs.~\eqref{eq:ENCorr}-\eqref{eq:EMCorr}, which also show that the total phase lag is the resultant of the signal transduction and transcription lags. For a signal with period $T=5 \text{ min}$ and a gene with degradation rate $\lambda =  0.04 \text{ min}^{-1}$~\cite{Corrigan:2014},  the transcription phase lag is $\arctan(\omega/\lambda) \simeq \pi/2$, which corresponds to a delay of $\simeq 1.25 \text{ min}$. Given a measured total mean lag of $9 \pi/10$, this implies that the signal transduction introduces a phase lag $\Delta \simeq 2 \pi / 5$, equivalent to a delay of~$\simeq 1 \text{ min}$.  }
\label{fig:bioinsightcartoon}
\end{figure} 

As a motivating example, we consider a recent experiment~\cite{Corrigan:2014} 
measuring single-cell time courses of the expression of gene \emph{csaA} in \textit{Dictyostelium discoideum} 
when driven by a naturally oscillating extracellular cAMP signal. 
Corrigan and Chubb found that whilst individual single-cell time traces displayed no clear entrainment, with considerable heterogeneity both across time and across the population, 
there was a clear correlation between the external cAMP phase (measured by proxy through the cell speed) 
and the population-averaged, time-dependent level of \emph{csaA} transcripts.  
This suggests that $\mean{N_t}$ could generate precision in cell choices at the population level~\cite{Corrigan:2014}.

The experiment showed that 
the population-averaged mRNA expression was approximately sinusoidal.
Hence, the data can be fitted to the function
\begin{equation}
\label{eq:ENCorr}
 \mean{N_t} = \beta \left [1 + \alpha\sin(\omega t) \right ].
\end{equation}
Assuming constant degradation rate $\lambda$, Eqs.~\eqref{eq:Xdiffeq}~and~\eqref{eq:ENXM} 
show that the upstream transcription rate is also sinusoidal with the same frequency, 
yet with a modified amplitude and a 
phase shift (Fig.~\ref{fig:bioinsightcartoon}):
\begin{align}
 \label{eq:EMCorr}
 \mean{M_t} = 
 \beta
 \left[\lambda + \alpha \sqrt{\omega^2 + \lambda^2}
 \sin\left(\omega t+\arctan \left(\frac{\omega}{\lambda} \right) \right)\right].
\end{align}
This is similar to phase relationships in electrical and electronic circuits.

Consistent with Eq.~\eqref{eq:EMCorr},the cAMP phase and $\mean{N_t}$ were measured experimentally 
to have the same frequency $\omega \approx 2\pi/5 \approx 1.26 \ \text{ min}^{-1}$~\cite{Corrigan:2014}.
The experiments also showed that 
$\mean{N_t}$ had a mean phase lag of $9\pi/10$ (equivalent to a delay of $\approx 2.25 \text{ min}$) 
after the cAMP signal. Using the degradation rate 
$\lambda = 0.04  \text{ min}^{-1}$~\cite{Muramoto:2012} for gene \textit{csaA}, it follows that
the transcriptional phase lag is $\arctan(\omega/\lambda) = 0.49 \pi$, 
and signal transduction introduces a phase lag $\Delta \simeq 9\pi/10 - \pi/2 = 2 \pi/5$, equivalent
to a transduction delay $\Delta / \omega \simeq 1 \text{ min}$.
Hence our results can be used to adjudicate the time scales linked to cAMP signal transduction
within the cell.

Our model also clarifies the effect of the degradation rate $\lambda$ and frequency 
$\omega$ in the observed responses. Given the mRNA population average
oscillating around a mean value $\beta$~\eqref{eq:ENCorr},
the (unobserved) transcription rate oscillates with the same frequency $\omega$ around a
value $\beta \lambda$ and amplitude scaled by 
$\sqrt{\omega^2 + \lambda^2}$. 
The transcriptional phase lag $\arctan(\omega/\lambda)$ 
is bounded between 0 (when $\omega/\lambda \to 0$) and $\pi/2$ (when $\omega/\lambda\to\infty$).
Hence, large degradation rates reduce the phase lag and the amplitude of the mRNA oscillations 
downstream (through the dimensionless factor $\omega/\lambda$), and
reduce the mean value of mRNA expression independently of~$\omega$.

A similar analysis for the correlation function $\mean{M_tM_s}$ can be achieved by 
solving Eq.~\eqref{eq:ENXM2} numerically for given data.

\begin{figure*}[htb!]
\centerline{\includegraphics[width=\textwidth]{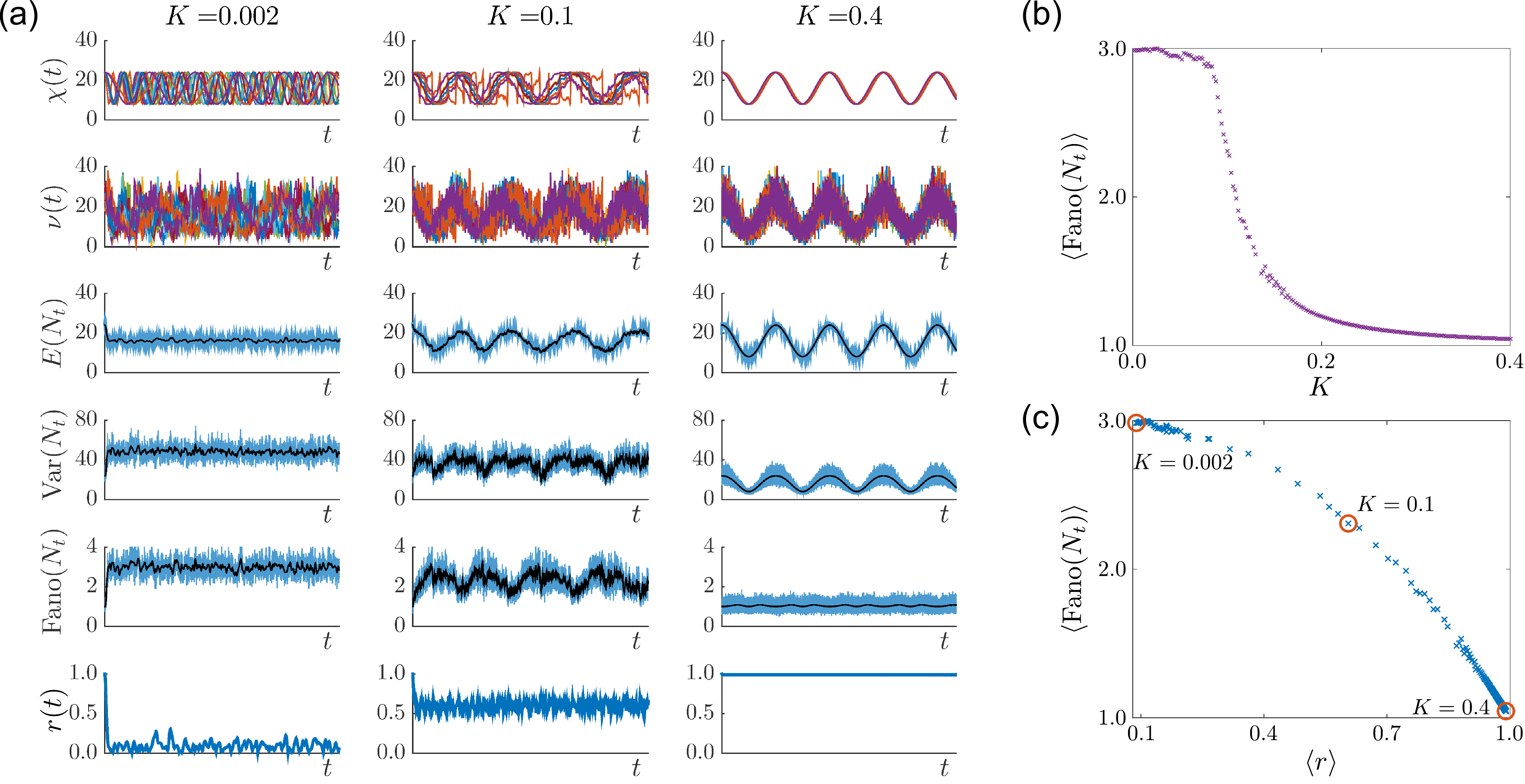}}
 \caption{\label{fig:kuramoto}
Noise characteristics of the Kuramoto promoter model~\eqref{eq:kuramoto}. 
(a) Numerical simulations for $C=100$ oscillatory cells and different 
coupling parameters: $K=0.002, 0.1, 0.4$ (left, middle, right columns).
For each coupling, the sample paths of the upstream effective drive $X$ and 
mRNA counts $N$ are shown. The mean, variance, and ensemble Fano factor of $N$
were calculated from the sample paths of $N$ (blue lines) and, more 
efficiently, from the sample paths of $X$ (black lines). The last row shows
the Kuramoto order parameter $r(t)$ measuring 
the cell synchrony, signaled by $r(t) \to 1$.
(b) Ensemble Fano factor (averaged over
the simulated time courses) against coupling parameter $K \in (0, 0.4]$. 
As $K$ is increased, the oscillators become synchronised and the
ensemble Fano factor decreases towards the Poisson value of unity.
(c) Scatter plot of the ensemble Fano factor against the order parameter $r(t)$ (both averaged
over the simulated time courses). As the oscillators
become synchronised  ($\langle r \rangle \to 1$), 
the ensemble Fano factor also approaches 1, signifying that the distribution
is Poissonian at all times. 
}
\end{figure*}

\subsection{Time-dependent ensemble Fano factor: a measure of synchrony in the 
population}
\label{sec:ensembleFano}

A commonly used measure of variability in the population is the 
\emph{ensemble Fano factor}: 
\begin{align}
\label{eq:Fano_def}
\R{Fano}(N_t)\defeq \frac{\R{Var}(N_t)}{\mean{N_t}},
\end{align} 
which is unity for the Poisson distribution. 
Its use has been popularised as a measure of the
deviation from the stationary solution of the transcription of an unregulated gene 
with constant rates~\cite{Thattai:2001, Raj:2009}, which is Poisson; hence 
with $\R{Fano}(N_t)\equiv 1, \, \forall t$. 

For time-varying systems, however, the ensemble Fano factor conveys 
how the dynamic variability in single cells combines 
at the population level. Indeed, $\R{Fano}(N_t)$ can be thought of as 
a measure of synchrony in the population at time $t$. 
For instance, it follows from Eq.~\eqref{eq:PoiResult} that the ensemble 
Fano factor for a population with perfectly synchronous drives 
is always equal to one, $\R{Fano}(N_t)\equiv 1, \, \forall t$.  
Even if the upstream drive $\chi(t)$ changes in time, 
the population remains synchronous and has a Poisson distribution at all times. 
On the other hand, under the assumptions of our model,
when $\R{Fano}(N_t)$ varies in time
it reflects a change in the degree of synchrony between cells, as captured
by the upstream drive $X_t$. 
Indeed, from~\eqref{eq:moments} it follows that:  
\begin{align*}
 \R{Fano}(N_t) & \defeq \frac{\R{Var}(N_t)}{\mean{N_t}} =
 \frac{\R{Var}(X_t)+\mean{X_t}}{\mean{X_t}} 
= 1 +  \R{Fano}(X_t).
\end{align*}
The greater the synchrony at time $t$, the closer $\R{Fano}(N_t)$ 
is to unity, since the deviation from the Poisson distribution emanates from
the ensemble Fano factor of the upstream drive $X_t$.

As an example, consider the Kuramoto promoter model
~\eqref{eq:kuramoto}-\eqref{eq:kuramoto2} 
introduced in Section~\ref{sec:KPM},
where the cells in the population become more synchronised 
as the value of the coupling $K$ is increased.
Figure~\ref{fig:kuramoto} shows simulation results for 100 cells
with a range of couplings. 
The order parameter $r(t) \in [0,1]$
measures the phase coherence of the oscillators at time $t$;
as it grows closer to 1, so grows the degree of synchrony. 
Using the Kuramoto numerics, we calculate the ensemble Fano factor 
$\R{Fano}(N_t)$ for the transcription model. 
As seen in Fig.~\ref{fig:kuramoto}(b)-(c),
the more synchronous,
the closer the Fano factor is to the Poisson value, i.e.,
 $\langle r (t) \rangle \to 1 \implies \langle \R{Fano}(N_t) \rangle \to 1$.

Figure~\ref{fig:kuramoto} also illustrates the computational
advantages of our method. The cost to approximate 
the time-varying ensemble moments is drastically reduced 
by using~\eqref{eq:moments}, because transcription and 
degradation events do not have to be simulated. 
The sample paths of the effective drive $\chi_i(t)$ 
were used to estimate the time-varying moments: 
$\mean{N_t} = \mean{X_t}$ and $\R{Var}({N_t})=\R{Var}(X_t)+\mean{X_t}$ 
(shown in black). These correspond to the numerical simulation of ODEs, and 
are far more efficient than sampling from realisations $\nu_i(t)$
of the mRNA copy number.

\section{Variability over time: stationarity and 
ergodicity}
\label{sec:stationarity}

Our results up to now have \textit{not} assumed
stationarity; in general, the distribution~\eqref{eq:PoiMixResult} 
and moments~\eqref{eq:moments} depend on time. 
If the cells in the population are uncorrelated
and both $M$ and $L$ are \emph{stationary} 
(i.e., their statistics do not change over time) 
then $f_{X_t}(x,t)$ tends to a stationary
density $f_{X_t}(x)$~\cite{Soong}, and 
the full solution $P(n,t)$ also tends to a stationary distribution $P(n)$. 

\begin{figure}[h]
\centerline{\includegraphics[width=0.5 \textwidth]{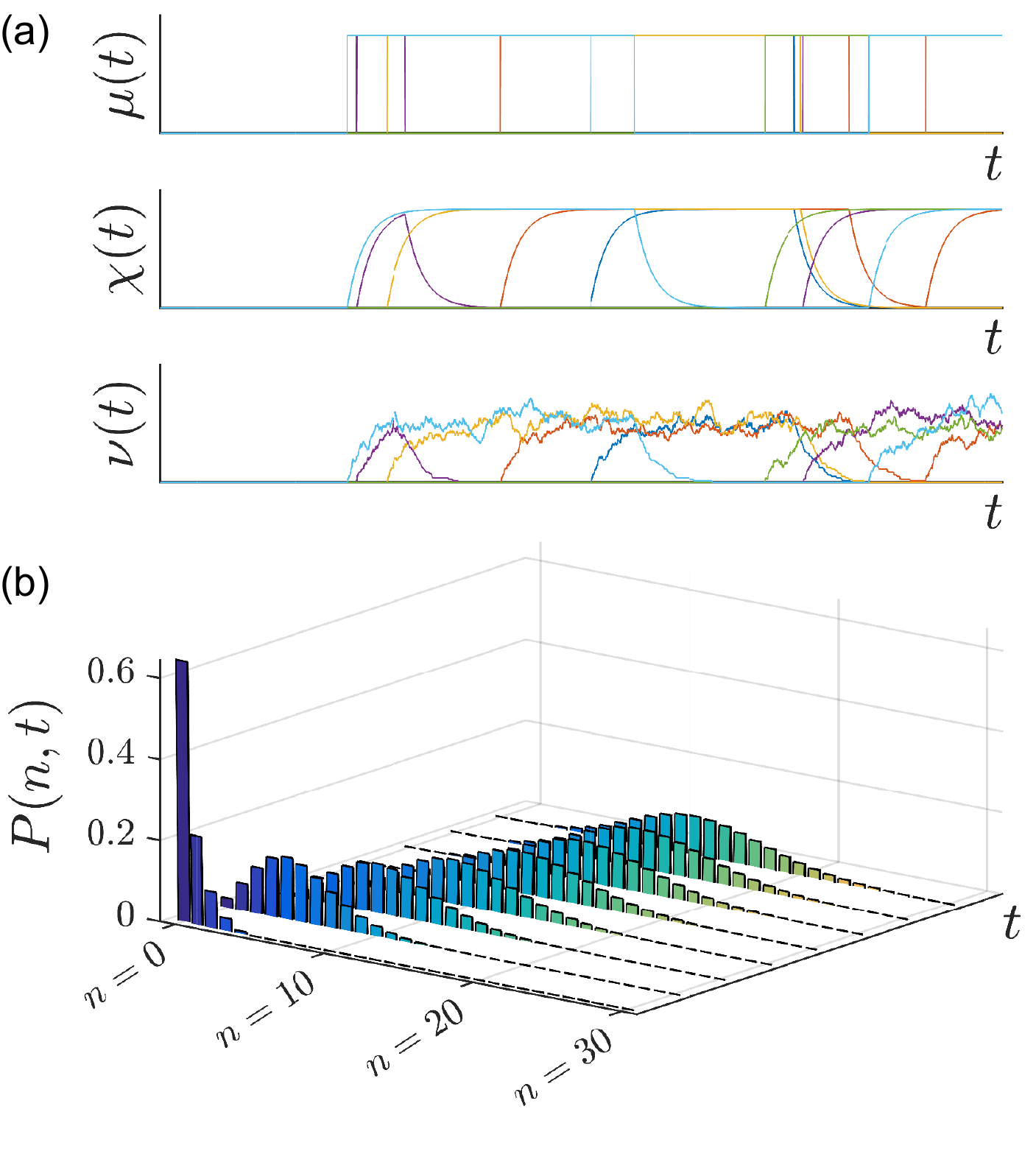}}
\caption{\label{fig:IB} `Burn-in' transient in the random telegraph (RT) model. 
(a) Sample paths of the transcription rate $M$, the effective upstream drive $X$, 
and the number of mRNAs $N$ for an initial condition $P(0,0)=1$ 
with all cells initialised in the inactive state~\cite{Iyer-Biswas:2009}.
(b)  The full solution of the RT model for this initial probability distribution 
exhibits an exponential decay as the system approaches its stationary distribution. 
The delta distribution at $t=0$ is omitted for scaling purposes.
}
\end{figure}

Under such assumptions leading to stationarity, any time dependence in $P(n,t)$ 
only describes the `burn-in' transient from an initial condition towards the
attracting stationary distribution, as discussed above in Section~\ref{sec:sync_deriv}.
Several examples of such transience have been studied in the literature, both 
in gene-state switching models with constant rate parameters~\cite{Iyer-Biswas:2009, Shahrezaei:2008, Vandecan:2013}, 
and in a model with state-dependent rates~\cite{Smith:2015}, 
to describe how the distribution $P(n,t)$ settles to stationarity when the process
is started from an initial Kronecker delta distribution $P(n,0)=\delta_{n0}$. Figure~\ref{fig:IB} and 
Appendix~\ref{app:IB} analyse this transience explicitly for the random telegraph model.

If, in addition to stationarity, we assume the cells to be indistinguishable, 
the population is \emph{ergodic}. In this case, the distribution obtained from a single cell 
over a time $T$, as $T \to \infty$, is equivalent to the distribution obtained from a 
time snapshot at stationarity of a population of $C$ cells, as $C \to \infty$, i.e.,
\begin{align}
& \hspace*{2.5in} P(n) =  \langle P(n) \rangle     \label{eq:ergodic_equiv} \\
&\text{where} \quad
P(n) \defeq \lim_{t \to \infty} P(n,t) 
= \lim_{t \to \infty} \lim_{C \to \infty} \frac{1}{C} \sum_{i=1}^C  
\frac{\chi_i(t)^n}{n!} e^{-\chi_i(t)} f_{X_t}(\chi_i(t),t)
= \int \frac{x^n}{n!}\,e^{-x} \,f_{X_t}(x) \,dx,   
\label{eq:Ccells} \\
&\text{and} \quad
\langle P(n) \rangle  \defeq  \lim_{T\to\infty}   \frac{1}{T}\int_0^T P(n,t | 
\chi(t))\D t  = \lim_{T\to\infty} \frac{1}{T}\int_0^T  
\frac{\chi(t)^n}{n!}e^{-\chi(t)} \D t.
\label{eq:ergodic}
\end{align}
Here, $\langle . \rangle$ denotes time-averaging,
and $\chi(t)$ in Eq.~\eqref{eq:ergodic} is the sample path of the effective drive for a randomly chosen cell.
Therefore, under the assumption of ergodicity, 
the averages computed over single-cell sample paths
can be used to estimate the stationary distribution of the population.

\subsection{Ergodic systems: stochastic \emph{vs} deterministic drives}

\begin{figure}[ht!]
 \centering
 \centerline{\includegraphics[width = .65 \textwidth]{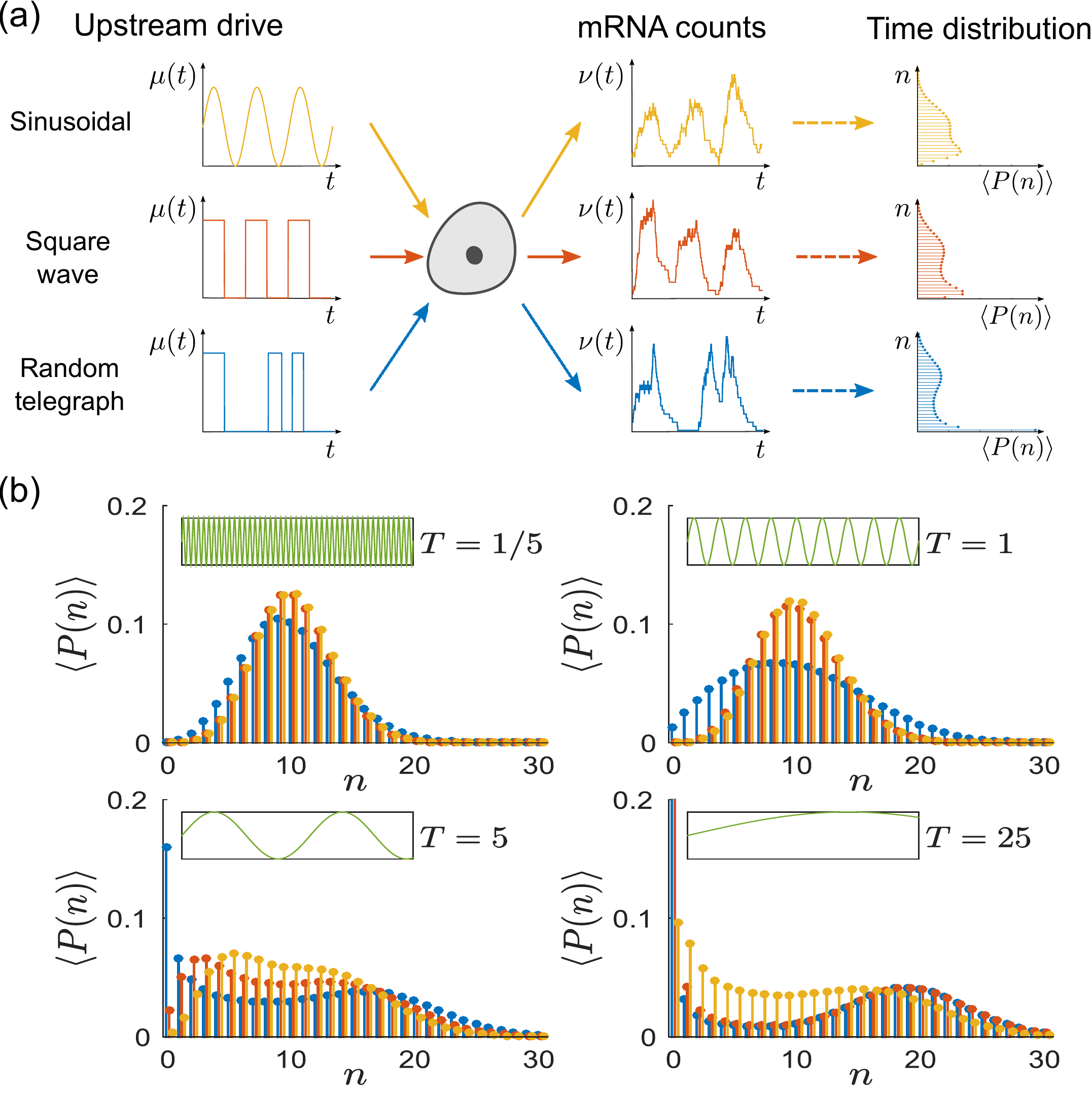}}
 \caption{\label{fig:stoch_det} Ergodic transcription models 
 under periodic and stochastic upstream drives. 
 (a) We consider gene transcription under three drives $\mu(t) \in [0,20]$: 
 a sinusoidal wave with period $T$  (yellow);  
a square wave with period $T$ (red); 
a random telegraph process with expected waiting time $T/2$ in each state 
(blue).
For such ergodic systems, the distribution computed over time $\langle P(n) \rangle$ 
corresponds to the stationary distribution.
(b) The distribution  $\langle P(n) \rangle = P(n)$ presents distinct features
as the period $T$ is varied. 
}
\end{figure}

It has been suggested that stochastic and periodic drives lead to distinct properties 
in the noise characteristics within a cell population~\cite{Hilfinger:2011}.
We investigate the effect of different temporal drives on the full distribution~\eqref{eq:PoiMixResult}
under ergodicity using~\eqref{eq:ergodic_equiv}-\eqref{eq:ergodic}. Note that 
when $\chi(t)$ is periodic
with period $T$, the limit in Eq.~\eqref{eq:ergodic} is not required.
In Figure~\ref{fig:stoch_det}, we show the time-averaged distribution 
$\langle P(n) \rangle$ for a cell under three different upstream drives $\mu(t)$: 
(i)~a continuous sinusoidal form; 
(ii)~a discontinuous square wave form; 
(iii)~a random telegraph (RT) form, which can
be thought of as the stochastic analogue of the square wave.
In all cases, the drive $\{\mu(t)\}_{t\ge 0} \in [0,20]$ 
with the same period, or expected period, $T$.
For simplicity, we set $\lambda(t) \equiv 1$.

As the period $T$ is varied, 
the similarity between the distributions under the three upstream
drives varies considerably
(Fig.~\ref{fig:stoch_det}).  At small $T$, 
the distributions under sinusoidal and square wave forms 
are most similar; whereas at large $T$, the distributions under 
square wave and RT forms become most similar.
In general, the distribution of the RT model has longer tails
(i.e., $n$ low and high) as a consequence of
long (random) waiting times that allow the system 
to reach equilibrium in the active and inactive states, 
although this effect is less pronounced 
when the promoter switching is fast 
relative to the time scales of transcription and degradation
(e.g., $T=1/5$).
On the other hand, as $T$ grows, the square wave and RT drives
are slow 
and the system is able to reach the equilibrium in both active and inactive 
states. 
Hence the probability distributions of the square wave and RT drives 
become similar, with a more prominent bimodality.

\subsection{The temporal Fano factor: windows of stationarity in single-cell time course 
data}
\label{sec:temporalFano}

The \emph{temporal Fano factor} (TFF) is defined similarly to the ensemble 
version~\eqref{eq:Fano_def}, 
but is calculated from the variance and mean of a \emph{single time series} 
$\{\nu(t)\}_{t\ge 0}$ over a time window $W\defeq (t_1, t_2)$: 
\begin{equation}
\label{eq:FanoT}
 \R{TFF}_{\{\nu(t)\}}(W)  \defeq 
 \frac{\langle \nu(t)^2 \rangle_{t\in W} - 
 \langle \nu(t) \rangle^2_{t\in W}}{\langle \nu(t) \rangle_{t\in W}}.
\end{equation}
In fact, this is the original definition of the Fano factor~\cite{Fano:1947}, which
is used in signal processing to estimate statistical fluctuations of a 
count variable over a time window.
Although $N_t$ is not a count variable
(it decreases with degradation events), 
the TFF can be used to detect windows of stationarity 
in single-cell time courses.

\begin{figure}[ht!]
\centerline{\includegraphics[width=.55\textwidth]{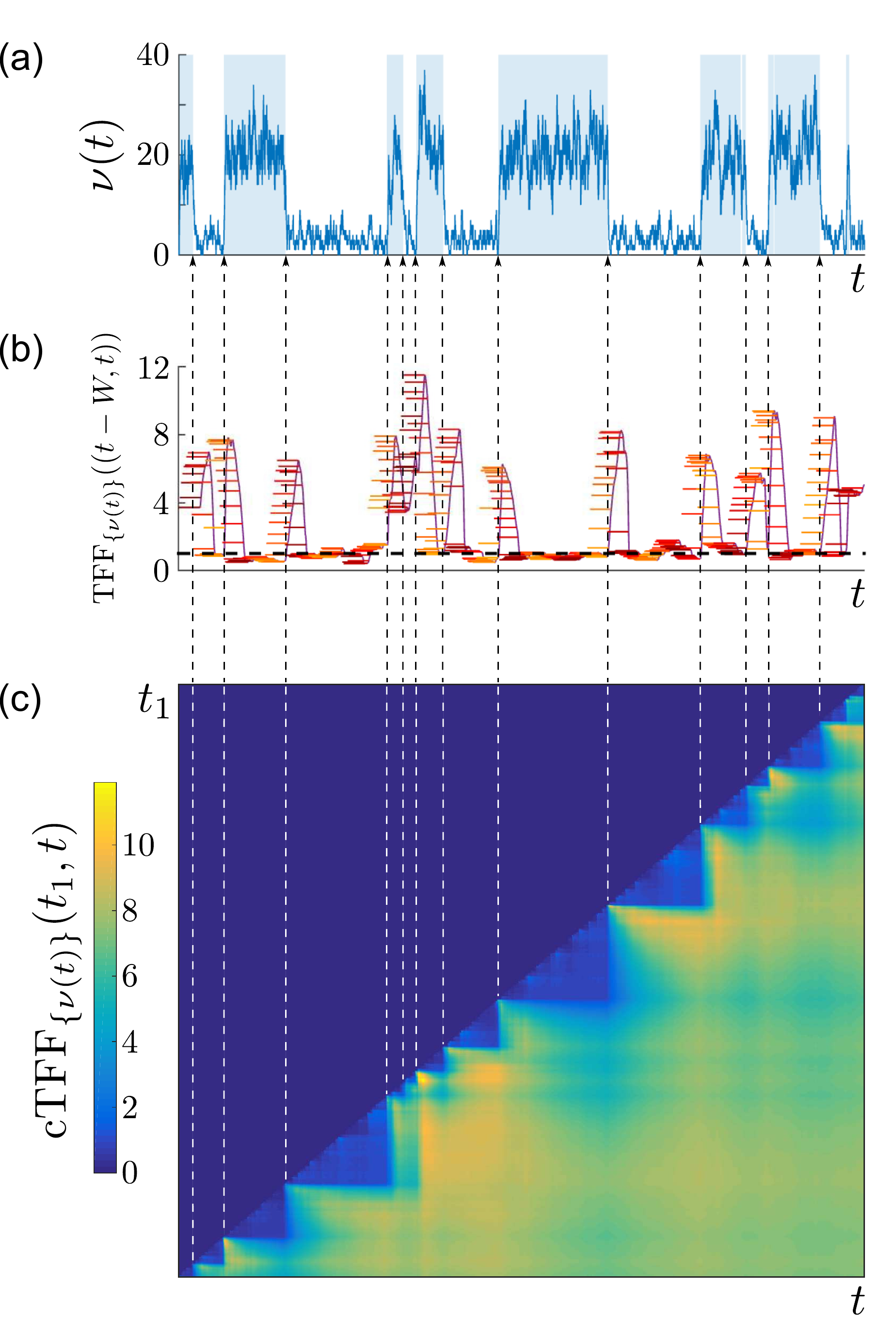}}
\caption{The temporal and cumulative Fano factor. 
(a) A sample path $\{\nu(t)\}_{t\ge 0}$ of mRNA counts
from the (leaky) RT model. The time periods when the gene is in the active state are shaded.
(b) The temporal Fano factor~\eqref{eq:FanoT},
 $\R{TFF}_{\{\nu(t)\}}((t-W,T))$, computed over a time window $W$ of fixed length 
 indicated by the horizontal bars at each $t$. 
 When $W$ extends over a stationary section of the sample path, TFF 
 is close to unity, corresponding to the Poisson distribution (black dashed line). 
(c) Heatmap of the cumulative Fano factor~\eqref{eq:cTFF},  $\R{cTFF}_{\{\nu(t)\}}(t_1, t)$,
defined only for $t \ge t_1$.
Note the marked step pattern corresponding to the switching times,
indicated by dashed lines as a guide to the eye.
}
\label{fig:tempfano} 
\end{figure}

Figure~\ref{fig:tempfano}a shows a single-cell sample path $\{\nu(t)\}_{t\ge0}$ 
generated by the (leaky) random telegraph model with constant 
degradation rate $\lambda$, and transcription rates $\mu_1 > \mu_0 > 0$
for the active and inactive promoter states.  
The leaky RT model is equivalent to the standard RT model, and
switches between two states with expectations $\mu_1/\lambda$ and 
$\mu_0/\lambda$.
In the time windows $W$ between promoter switching, 
$\{\nu(t)\}_{t \in W}$ can be considered almost at stationarity 
and described by a Poisson distribution with parameter 
$\mu_0/\lambda$ (resp. $\mu_1/\lambda$) in the 
inactive (resp. active) state.
Hence $\R{TFF}_{\{\nu(t)\}}(W) \simeq 1 $ across most of
the sample path, except over the short transients $W_{\text{trans}}$ 
when the system is switching between states, where 
$\R{TFF}_{\{\nu(t)\}}(W_\text{trans}) > 1 $ (Fig.~\ref{fig:tempfano}b). 

Alternatively, this information can be extracted robustly from 
the \emph{cumulative Fano factor} (cTFF):
\begin{equation}
\label{eq:cTFF}
 \R{cTFF}_{\{\nu(t)\}}(t_1, t)  =  \R{TFF}_{\{\nu(t)\}}\left( (t_1, t) \right ),\quad   t \ge t_1
\end{equation}
which is computed over a \emph{growing window} from a fixed starting time 
$t_1$.
The cTFF is a cumulative moving average giving an integrated description 
of how the stationary regimes are attained between switching events
indicated by the step-like structure of the heatmap in
Fig.~\ref{fig:tempfano}c.

\section{Discussion}

We have presented the solution of the master equation for 
gene transcription with upstream dynamical variability 
in a setting that allows a unified treatment of a broad class of models,
enabling quantitative biologists to go beyond  
stationary solutions when analysing noise sources in single-cell experiments. 
As a complementary approach to the explicit stochastic simulation of 
networks with many genes to account for the variability in data, 
our work uses a parsimonious transcription model of Poissonian type that 
includes explicitly the effect of dynamical and cell-to-cell upstream variability 
in the master equation.
We show that the solution to this gene transcription-degradation model can be expressed 
as a Poisson mixture form~\eqref{eq:PoiMixResult}.
This solution can be interpreted as the combination of 
an upstream component (dynamic or static; deterministic or stochastic)
with a downstream Poissonian immigration-death process.
Since only the upstream process is model-specific, 
different models are solved by obtaining the different mixing 
densities $f_{X_t}$ of the upstream process. 
This generic mathematical structure can describe both time-dependent
snapshots across the population, as well as the dynamical variability over
single-cell time courses in a coherent fashion.

The solution~\eqref{eq:PoiMixResult} can also be understood from the perspective of 
Gardiner and Chaturvedi's  \textit{Poisson representation}~\cite{Gardiner:1977,Gardiner}.
Originally, the Poisson representation was introduced as an \textit{ansatz} 
for asymptotic expansions of 
stationary systems, and included only constant rate parameters. 
Hence the original Poisson representation \textit{ansatz} has not been used 
widely for time-varying solutions~\cite{Gardiner:1977}.
In contrast, our time-dependent Poisson mixture~\eqref{eq:PoiMixResult} 
is obtained here as a solution to a
non-stationary ME model, rather than backwards via basis expansions,
and the mixing density $f_{X_t}$ 
has a physical interpretation in terms of single-cell sample paths relatable to data.

In this respect, our preparatory result~\eqref{eq:MESolution} for the perfectly synchronous population
can be thought of as an extension of the `Poisson representation' to include time-varying rate parameters.
Note also that this perfectly synchronous solution~\eqref{eq:MESolution} corresponds to a particular case of the
multi-gene solution obtained by Jahnke and Huisinga~\cite{Jahnke:2007}.
However, \eqref{eq:MESolution} does not yet encapsulate the cell-to-cell variability.
It is the full Poisson mixture solution~\eqref{eq:PoiMixResult} that extends 
the scope of the time-varying `Poisson representation' a step further, by allowing 
for stochastic rate parameters that can describe cell-to-cell variability as well as dynamic variability.

\begin{figure}[t!]
\vspace*{4mm}
\centerline{\includegraphics[width=.75\textwidth]{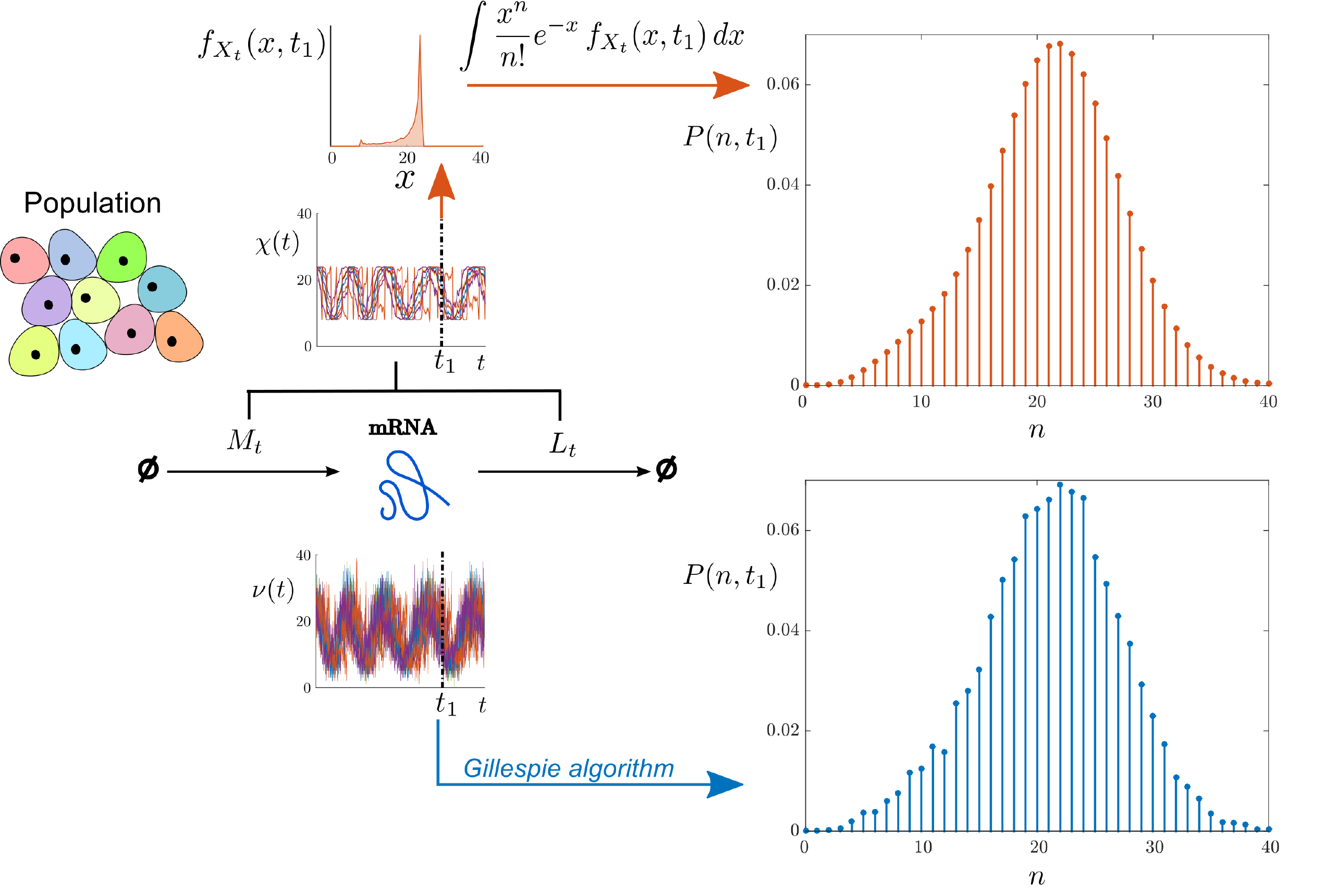}}
\caption{
Efficient sampling of the full distribution 
$P(n,t)$ for transcription with upstream cellular drives.
We consider upstream drives governed by the Kuramoto promoter model~\eqref{eq:kuramoto}
for $C=10,000$ coupled oscillatory cells.  
Sample paths of $N$ are simulated directly with the Gillespie algorithm 
to approximate $P(n,t)$ at time $t_1$ (bottom, blue). 
Alternatively, sample paths of $X$ are used to estimate
 $f_{X_t}$, which is then mixed by performing the 
 numerical integration~\eqref{eq:PoiMixResult} to obtain $P(n,t)$ (top, red). 
The latter sampling through $X$ is more regular and far less costly: 
CPU time via $N$ is $\approx 36000\, s$ whereas CPU time via $X$ 
is~$\approx 0.1 \,s$.}
\label{fig:cost_reduction} 
\end{figure}

Our solution confers two broad advantages. 
The first is pragmatic: 
Since $X_t$ is a continuous random variable 
satisfying a linear random differential equation, we can draw upon the 
rich theory and analytical results for $f_{X_t}$, even for non-stationary models,
or we can use ODE and PDE solvers as further options to solve 
the differential equation for $f_{X_t}$. 
If simulations are still necessary, sampling
$P(n,t | M,L)$ directly using stochastic simulation algorithms becomes computationally
expensive, particularly if the upstream processes $M$ and $L$
are time-varying~\cite{Shahrezaei:2008b}.
Instead, we can sample $f_{X_t}(x,t)$ directly using the 
random differential equation~\eqref{eq:Xdiffeq},
and then obtain the full distribution via numerical integration 
using~\eqref{eq:PoiMixResult}. This approach leads
to a significant reduction in computational cost, as 
shown in Fig.~\ref{fig:cost_reduction}.

Our approach can also be used to analyse noise characteristics 
in conjunction with biological hypotheses. 
If measurements of additional cellular variables (e.g., 
cell cycle) are available, they can be incorporated as a source of variability 
for gene regulation to test biological hypotheses computationally 
against experimental data. Conversely, it is possible to discount
the Poissonian component from observed data, so as to fit different promoter models 
to experimental data and perform model comparison~\cite{thesis}. Our discussion
of a recent experiment of gene expression driven by cAMP signalling~\cite{Corrigan:2014}
exemplifies this approach.

The second advantage of our framework is conceptual. 
Through the natural decoupling of the solution into 
a discrete, Poisson component (`downstream') and a continuous, mixing component (`upstream'), 
we derive time-dependent expressions for both ensemble and temporal 
moments, recasting the concept of `intrinsic'/`extrinsic' noise
for dynamic upstream cellular drives. 
Importantly, all upstream variability gets effectively imbricated 
through the upstream effective drive $X$, which can be
interpreted in terms of a biochemical differential rate equation.
This analysis clarifies how upstream fluctuations
are combined to affect the probability distribution of the mRNA copy number, 
providing further intuition about the sources of noise and their temporal characteristics.  
Indeed, stripping the model down to its 
extrinsic component $f_{X_t}$ can provide us with additional 
understanding of its structure and time scales~\cite{thesis}. 

Finally, although we have concentrated here on the amenable analytical solutions that can be obtained for 
the single gene case, we remark that our solution could be extended to monomolecular multi-gene networks,
by merging Jahnke and Huisinga's result~\cite{Jahnke:2007} for synchronous networks
with our mixture result for the asynchronous case. Such a generalisation could be implemented
computationally to reduce the cost of simulating stochastic networks. 
Such an extension will be the subject of future work. 
The solutions of higher order reaction systems obtained through the Poisson 
representation \textit{ansatz} could also be extended to include stochastic 
rates.  This approach could lead to deeper understanding of 
models with and without feedback~\cite{Iyer-Biswas:2014}.

\section*{Acknowledgments}
We are grateful for insightful comments and extended discussions with Juan 
Kuntz, Philipp Thomas, Martin Hemberg and Andrew Parry. 
JD was funded by an EPSRC PhD studentship. 
MB acknowledges funding from the EPSRC through grants EP/I017267/1 and
EP/N014529/1.

\paragraph*{Authors' contributions:}
JD carried out the calculations and simulations, participated in the design 
of the study, and wrote the manuscript. MB conceived of the study, 
designed the study, coordinated the study, and wrote the manuscript. Both 
authors gave final approval for publication.

\paragraph*{Data statement:}  No new data was collected in the course of this research.

\bibliographystyle{abbrv}
\bibliography{refs}

\clearpage

\appendix
\section*{Appendices}

\section{The `burn-in' transient towards stationarity}
\label{app:N_ic}

In Section~\ref{sec:sync_deriv}, it was stated that 
the contribution from the initial condition $N_t^{ic}$ decreases exponentially 
for biophysically realistic degradation rates $\{\lambda (t)\}_{t \geq 0}$. 
As a result, the transcripts that were present at $t=0$ are expected to degrade 
in finite time, 
and the long-term population is expected to be composed only of mRNA molecules 
that were transcribed since $t=0$.

Let the initial condition be described by a random variable $N_0$ with a 
given probability
distribution. It follows from Eqs~\eqref{eq:MESolution}-\eqref{eq:PnPn0mix}
that
\begin{align*}
P \left (n,t \, | \, \{\mu(\tau)\}_{\tau\in [0,t]}, 
\{\lambda(\tau)\}_{\tau\in[0,t]} \right) 
& =
\sum_{n_0} P\left(n,t  \, | \, \{\mu(\tau)\}_{\tau\in [0,t]}, 
\{\lambda(\tau)\}_{\tau\in [0,t]}, n_0 
\right) \,\R{Pr}(N_0=n_0) \\
&= \sum_{n_0} \sum_{k=0}^n \R{Pr}(N_t^{ic}=k|N_0=n_0)\,\R{Pr}(N_t^s=n-k) 
\,\R{Pr}(N_0=n_0) \\
& = \sum_{k=0}^n \R{Pr}(N_t^s=n-k) \left [\sum_{n_0} \R{Pr}(N_t^{ic}=k|N_0= 
n_0)\,\R{Pr}(N_0=n_0) \right] \\
&=  \sum_{k=0}^n \R{Pr}(N_t^s=n-k) \, \R{Pr}(N_t^{ic}=k),
\end{align*}
where $N^{ic}_t|n_0$ and $N^s_t$ are distributed according to 
Eqs~\eqref{eq:NicNs}-\eqref{eq:Ns},
and $\R{Pr}(N_t^{ic}=k)$
is the mixture of the time-dependent binomial 
distribution~\eqref{eq:NicNs} and the distribution of the initial condition 
$N_0$. 

\subsection{`Burn-in' transience in the model with constant transcription and degradation}
\subsubsection*{The decay towards stationarity} 
To understand the `burn-in' period more explicitly, consider the simplest example of the gene 
transcription model~\eqref{reaction_diagram} with constant transcription and 
degradation rates $\mu$ and $\lambda$, 
and assume that there are initially $n_0$ mRNA transcripts. Given that $N_0 \sim 
\delta_{0,n_0}$,
the solution is given by:
\begin{align}
 N_t &= N_t^{ic}|n_0+N_t^s \quad \text{ where } \nonumber\\
 N_t^{ic}|n_0 & \sim \R{Bin}\left(n_0,e^{-\lambda t}\right), \nonumber\\
 N_t^s & \sim \R{Poi}\left(\frac{\mu}{\lambda} \left(1-e^{-\lambda t} \right) 
\right). \label{eq:poi_constant}
\end{align} 
Hence 
as $t\to\infty$, the distribution will tend towards 
$\R{Poi}(\mu/\lambda)$, the stationary distribution of the population. 
This is a well-known result in the literature~\cite{Gardiner, Bremaud}.

\subsubsection*{Starting at stationarity: the time-dependent $N_t^{ic}$ and 
$N_t^s$ balance each other at all times}
\label{ex:NicNsbalance}

It is illustrative to consider the dynamics of this system when
the initial condition is chosen to be the stationary distribution.
In this case,
the  breakdown of $N_t$ into the 
time-dependent components $N_t^{ic}$ and $N_t^s$ 
will need to reproduce the stationary distribution at all times $t>0$, 
with no `burn-in' period.

To see this, let the initial distribution start at stationarity, i.e.,
$ N_0 \sim \R{Poi}(\mu/\lambda)$ and
\begin{equation*}
 \R{Pr}(N_0=n_0) = \left(\frac{\mu}{\lambda}\right)^{n_0}
\frac{e^{-\mu/\lambda}}{n_0!}.
\end{equation*}
The distribution of $N_t^s$ is still given by~\eqref{eq:poi_constant}
and the contribution of $N_t^{ic}$ is given by 
\begin{align*}
 &\R{Pr}(N_t^{ic}=k) 
 = \sum_{n_0=k}^{\infty}\R{Pr}(N_t^{ic}=k|N_0=n_0) \,\R{Pr}(N_0=n_0) \\
 &= \sum_{n_0=k}^{\infty} \binom{n_0}{k} \left(e^{-\lambda t}\right)^k
 \left(1-e^{-\lambda t}\right)^{n_0-k}\,
 \left(\frac{\mu}{\lambda}\right)^{n_0}\frac{e^{-\mu/\lambda}}{n_0!} 
 = \left(e^{-\lambda t}\right)^k\, e^{-\mu/\lambda}
 \sum_{r=0}^{\infty} \left(\frac{\mu}{\lambda}\right)^{r+k}
 \frac{\left(1-e^{-\lambda t}\right)^r}{k!\,r!} \\
 &= \left(\frac{\mu}{\lambda}\right)^k 
 \frac{\left(e^{-\lambda t}\right)^k\,e^{-\mu/\lambda}}{k!}
 \sum_{r=0}^{\infty} \left(\frac{\mu}{\lambda}\right)^r 
 \frac{\left(1-e^{-\lambda t}\right)^r}{r!} 
 = \left(\frac{\mu}{\lambda}\right)^k 
 \frac{\left(e^{-\lambda t}\right)^k\,e^{-\mu/\lambda}}{k!}\,
 e^{\frac{\mu}{\lambda}(1-e^{-\lambda t})} = \left(\frac{\mu}{\lambda}\, e^{-\lambda t}\right)^k 
 \frac{e^{-\frac{\mu}{\lambda}e^{-\lambda t}}}{k!}.
\end{align*}
In other words, $N_t^{ic}\sim\R{Poi}\left(\frac{\mu}{\lambda}\,e^{-\lambda 
t}\right)$, which cancels the contribution from 
$N_t^s\sim\R{Poi}\left(\frac{\mu}{\lambda}\left[1-e^{-\lambda 
t}\right]\right)$. 
Therefore
\begin{equation*}
 N_t\sim\R{Poi}\left(\frac{\mu}{\lambda}\right), \quad \forall t.
\end{equation*}
This example shows that $N_t^{ic}$ and $N_t^s$ will combine to reproduce a
stationary distribution at all times $t>0$, when the system starts at 
stationarity so that there is no `burn-in' transient.

\subsubsection*{Starting the system at $t = - \infty$}
The same is true if the system is not stationary but
we start the system at $t=-\infty$ with any initial condition. 
Then, for $t>0$, the system will be independent of the initial condition 
and will be described by $N_t^s$.

Let us denote the state of the system for $t>0$ by 
the \emph{attracting} distribution $P_*$. Although 
$\R{Pr}(N_t^s=n)=P_*(n,t), \forall n \in\mathbb{N}, \forall t>0$, 
we wish to distinguish $P_*$ from 
$P_s$ because we only have equality of the two distributions when the system 
starts at $t=-\infty$. $P_*$ can be thought of as an inherent property of the 
system, analogous to the stable point of a dynamical system that moves in 
time (sometimes called a \emph{chronotaxic system}~\cite{Suprunenko:2013}).

If $\R{Pr}(N_0=n_0)=P_*(n_0,0)$ for all $n_0$ in 
Eq.~\eqref{eq:PnPn0mix}, the 
contributions from $N_t^{ic}$ and $N_t^s$ balance each other as they did in the
case of stationarity with $N_0 \sim \R{Poi}(\mu/\lambda)$,
and we would have $P(n,t) = P_*(n,t), \forall n \in\mathbb{N}, 
\forall t>0$ (recall that the breakdown $N_t = N_t^{ic} + N_t^s$ simply resolves
the existing mRNA molecules into those that were present at $t=0$, and those that
were transcribed since $t=0$). 
Thus we only observe an initial transient period if the initial 
distribution starts away from its attracting distribution at $t=0$.
In all other cases, the following mathematical formulations are equivalent: 
\emph{i)} assume that the system was initialised at $t=-\infty$ and consider 
only $N_t^s$, or \emph{ii)} use the initial distribution $P_*(n,0)$ for all $n$ 
at $t=0$, and consider $N_t^{ic} + N_t^s$.

In this work, we focus on the time dependence of $P(n,t)$ 
induced through non-stationarity of the parameters, and/or synchronous 
behaviour of the cells within the population. Hence, unless otherwise stated, 
in this work we assume that the system was initialised at $t=-\infty$ and that the distribution 
of $N_t^s$ is the attracting distribution $P_*(n,t)$ for all $t>0$, i.e., we neglect the contribution from $N_t^{ic}$.

\subsection{`Burn-in' transience in the random telegraph model}
\label{app:IB}

A time-dependent solution of the probability generating function for the 
random telegraph model appeared in Ref.~\cite{Iyer-Biswas:2009}, although the 
explicit expression for $P(n,t)$ was omitted.  As discussed above, the RT model represents asynchronous
and stationary behaviour, hence the time-dependence appears only through 
convergence to the stationary distribution from the initial condition. We 
include the derivation here for 
completeness and to complement Fig.~\ref{fig:IB}.

Consider the RT model depicted in Fig.~\ref{fig:cycles}a.
Assuming that every cell is initialised in the inactive state with $n_0=0$ 
mRNA molecules, the probability generating function for the cell 
population is~\cite{Iyer-Biswas:2009}:
\begin{align*}
 &G(z,t) = F_s(z,t) \Phi_s(z,t) 
 - \frac{\mu \, \kn 
e^{-(\kn+\kf)t}}{(\kn+\kf)(1-\kn-\kf)} \, (1-z)   F_{ns}(z,t)\Phi_{ns}(z,t)
\end{align*} 
where $F_s(z,t) \defeq \tensor*[_1]{F}{_1}\left[-\kn,1-\kn-\kf;\mu e^{-t}(1-z) \right]$,
$\Phi_s(z,t) \defeq \tensor*[_1]{F}{_1}\left[\kn,\kn+\kf;-\mu(1-z) \right]$, 
$F_{ns}(z,t) \defeq \tensor*[_1]{F}{_1}\left[\kf,1+\kn+\kf;\mu e^{-t}(1-z) \right]$, and
$\Phi_{ns}(z,t) \defeq \tensor*[_1]{F}{_1}\left[1-\kf,2-\kn-\kf;-\mu(1-z) \right]$.
Here $\tensor*[_1]{F}{_1}\left(a,b;z\right)$ is the confluent 
hypergeometric function~\cite{Abramowitz}. 

Using the general Leibniz rule for differentiation, and omitting other details of the differentiation, we obtain
\begin{align*}
P(n,t) &= \frac{1}{n!}\frac{\partial^n G}{\partial z^n}\Bigg|_{z=0} \\ 
&= \frac{\mu^n}{n!} \sum_{r=0}^n \binom{n}{r} \frac{(-1)^r(-\kn)_r (\kn)_{n-r}\, e^{-rt}}{(1-\kn-\kf)_r(\kn+\kf)_{n-r}}\;
\tensor*[_1]{F}{_1}\left[-\kn+r,1-\kn-\kf+r;\mu e^{-t} \right] \\
&\hspace{7cm}\times\tensor*[_1]{F}{_1}\left[\kn+n-r,\kn+\kf+n-r;-\mu \right]\\
& + \frac{\kn\mu^{n+1} e^{-(\kn+\kf)t}}{(\kn+\kf)(1-\kn-\kf)n!}\sum_{r=0}^n \binom{n}{r} 
\frac{(-1)^r(-\kf)_r (1-\kf)_{n-r} e^{-rt}}{(1+\kn+\kf)_r(2-\kn-\kf)_{n-r}}\\
&\qquad\times\tensor*[_1]{F}{_1}\left[\kf+r,1+\kn+\kf+r;\mu e^{-t} \right]\;
\tensor*[_1]{F}{_1}\left[1-\kf+n-r,2-\kn-\kf+n-r;-\mu \right]\\
&-\frac{\kn\mu^n e^{-(\kn+\kf)t}}{(\kn+\kf)(1-\kn-\kf)n!}\sum_{r=0}^{n-1} \binom{n-1}{r} 
\frac{(-1)^r(-\kf)_r (1-\kf)_{n-1-r}\, e^{-rt}}{(1+\kn+\kf)_r(2-\kn-\kf)_{n-1-r}} \\
&\qquad\times\tensor*[_1]{F}{_1}\left[\kf+r,1+\kn+\kf+r;\mu e^{-t} \right]\; 
\tensor*[_1]{F}{_1}\left[-\kf+n-r,1-\kn-\kf+n-r;-\mu \right],
\end{align*}
where $(a)_m\defeq a(a+1)\dots(a+m-1)$ is Pochhammer's function~\cite{Abramowitz}. 

As $\zeta\to 0$, $\tensor*[_1]{F}{_1}\left[a,b;\zeta \right]\to1$. Hence 
as $t\to\infty$, $P(n,t)\to P(n)$, and we recover the known stationary solution:
\begin{equation*}
 P(n)= \frac{\mu^n}{n!} \frac{(\kn)_n}{(\kn+\kf)_n}\;
 \tensor*[_1]{F}{_1}\left[\kn+n,\kn+\kf+n;-\mu \right].
\end{equation*}

\section{The stationary solution for the refractory promoter model (3 cyclic promoter states)}
\label{app:3state}
As explained in Section~\ref{sec:3state}, the stationary solution of the
3-state cyclic model describing a refractory promoter is obtained by solving 
the set of equations:
\begin{align}
\label{eq:3state_equationson}
\frac{d}{d z}\left[(1-z)f_*(z)\right] &= 
-k_*f_*(z) + k_2f_2(z) \\
\label{eq:3state_equations1}
 \frac{d}{d z}\left[-zf_1(z)\right] &= -k_1f_1(z) + 
k_*f_*(z) \\
\label{eq:3state_equations2}
\frac{d}{d z}\left[-zf_2(z)\right] &= -k_2f_2(z) + 
k_1f_1(z)
\end{align}
to obtain an expression for $f_Z=f_*+f_1+f_2$.

Note that the transition matrix $[k_{sr}]$ containing the kinetic constants on the 
right-hand side of 
Eqs~\eqref{eq:3state_equationson}-\eqref{eq:3state_equations2}
is singular and hence $\lambda=0$ is an eigenvalue. Furthermore, by 
Gershgorin's circle theorem the non-zero eigenvalues of $[k_{sr}]$ have negative real 
parts, so a stationary solution always exists, i.e., the probabilities $\pi_i=\int_x f_i(x) 
\D x$ 
of being in state $S_i$ evolve to an equilibrium state given by 
the eigenvector $\bm{\hat{\pi}}$ associated with the eigenvalue 
$\lambda=0$. Note that $\bm{\hat{\pi}}$ must be normalised so that the elements 
sum to $1$. It can easily be shown that
\begin{equation}
\label{eq:3state_pi}
 \pi_i=\int_0^1 f_i(z)\D 
z=\frac{k_1k_2k_*}{k_i(k_2k_*+k_1k_*+k_1k_2)}.
\end{equation}
Now, integrating 
Eqs~\eqref{eq:3state_equationson}-\eqref{eq:3state_equations2} and 
using Eq.~\eqref{eq:3state_pi} we obtain the boundary values 
$f_1(1)=f_2(1)=0$ and $f_*(0)=0$. Also, summing 
Eqs~\eqref{eq:3state_equationson}-\eqref{eq:3state_equations2} and 
integrating gives
\begin{equation}
\label{eq:fisumC}
 -zf_1(z)-zf_{2}(z)+(1-z)f_*(z)=C,
\end{equation}
where $C$ is a constant.
Eq.~\eqref{eq:fisumC} is true for all $z\in[0,1]$, so we can substitute in 
$z=0$ or $z=1$ and use the fact that $f_1(1)=0$ and $f_1(1)=0$, or 
$f_*(0)=0$, to show that $C=0$. Hence 
\begin{equation}
\label{eq:fisum0}
 f_Z(z) = f_1(z)+f_2(z)+f_*(z) = \frac{1}{z}f_*(z),
\end{equation}
so we need only 
solve Eqs~\eqref{eq:3state_equationson}-\eqref{eq:3state_equations2} for 
$f_*(z)$, the marginal 
probability density corresponding to the active state.
Using~\eqref{eq:fisum0} and substituting 
into~\eqref{eq:3state_equationson}-\eqref{eq:3state_equations2}, we then obtain 
the 
following equation for $f_*(z)$:
\begin{align}
\label{eq:fon_ode}
0 &= z^2(1-z)f^{\prime\prime}_{*}(z)
 + z\left[(-3+k_1+k_2+k_*)z + 1-k_1-k_2 \right]f^{\prime}_{*}(z) \nonumber \\
&+ \left[(-1+k_1+k_2 - k_1k_2 -k_1k_*)z + k_1k_2 \right]f_*(z).
\end{align}
Set $f(z)=z^cu(z)$, where $c$ is a constant that we can choose, to 
transform~\eqref{eq:fon_ode} into 
\begin{align*}
 0&=z^{c+2}(1-z)u^{\prime\prime}(z) + z^{c+1}\left[(-3+\mathcal{K}-2c)z+1-k_1-k_2+2c \right]u^{\prime}(z) \\
 &+ z^c\left[(-1+\mathcal{K}-\tilde{\mathcal{K}}+(-2+\mathcal{K}-c)c)z\right.
\left. +k_1k_2-(k_1+k_2)c +c^2\right]u(z)
\end{align*}
where $\mathcal{K}:=k_1+k_2+k_*$ and 
$\tilde{\mathcal{K}}:=k_1k_2+k_1k_*+k_2k_*$.
From here, we set the last term on the right hand side to zero by choosing 
$c=k_1$ or $c=k_2$. We can then divide through by $z^{c+1}$ to obtain an 
equation in the form of the hypergeometric equation~\cite{Abramowitz}. 
For example, for $c=k_1$ we obtain
\begin{align*}
 0&=z(1-z)u^{\prime\prime}(z) 
 +\left[(-3-k_1+k_2+k_*)z+1+k_1-k_2\right]u^{\prime}(z) 
 +\left[-1-k_1+k_2+k_*-k_2k_*\right]u(z).
\end{align*}
When none of $\sqrt{(k_1-k_2-k_*)^2-4k_2k_*}$, $k_1-k_2$, or 
$k_*$ are integers, we can write down the solution~\cite{Abramowitz}:
\begin{align*}
 u(z) &= C_1\;\:\tensor*[_2]{F}{_1}\left[a^{(1)}_+,a^{(1)}_-;1+k_1-k_2; z\right]
+ C_2 \; z^{-k_1+k_2} \;\; \tensor*[_2]{F}{_1}\left[a^{(2)}_+,a^{(2)}_-;1-k_1+k_2; 
z\right]
\end{align*}
where $C_1$ and $C_2$ are constants of integration, $\tensor*[_2]{F}{_1}\left[a,b;c; z\right]$
is the Gauss hypergeometric function~\cite{Abramowitz}, 
and
\begin{align*} 
a^{(1)}_{\pm}&=\frac{1}{2}\left(2+k_1-k_2-k_*
\pm\sqrt{(k_*-k_1-k_2)^2-4k_1k_2} \right) \\ 
a^{(2)}_{\pm}&=\frac{1}{2}\left(2-k_1+k_2-k_*
\pm\sqrt{(k_*-k_1-k_2)^2-4k_1k_2} \right).
\end{align*}
The solutions for the other cases are similar and are also given 
in~\cite{Abramowitz}. 
Hence $f_*(z)=z^{k_1}u(z)$ is given by
\begin{align*}
 f_*(z) &= 
C_1\;z^{k_1}\;\tensor*[_2]{F}{_1}\left[a^{(1)}_+,a^{(1)}_-;1+k_1-k_2; z\right] 
+ C_2\;z^{k_2}\;\tensor*[_2]{F}{_1}\left[a^{(2)}_+,a^{(2)}_-;1-k_1+k_2; 
z\right].
\end{align*}
The same expression for $f_*(z)$ is obtained if we choose $c=k_2$ 
instead, so finally we can write down the general solution for $f_Z(z) = 
f_*(z)/z$:
\begin{align}
f_Z(z) &=C_1\;z^{k_1-1}\;\tensor*[_2]{F}{_1}\left[a^{(1)}_+,a^{(1)}_-;1+k_1-k_2; 
z\right] 
+ C_2\;z^{k_2-1}\;\tensor*[_2]{F}{_1}\left[a^{(2)}_+,a^{(2)}_-;1-k_1+k_2; 
z\right] \nonumber\\
&= C_1 \;z^{k_1-1}(1-z)^{k_*-1} \,\,  \tensor*[_2]{F}{_1}\left[a^{(1)}_+ + k_*-1,a^{(1)}_- 
+k_*-1;1+k_1-k_2; z\right] \nonumber\\
&+ C_2 \;z^{k_2-1}(1-z)^{k_*-1}  \,\,  \tensor*[_2]{F}{_1}\left[a^{(2)}_+ + k_*-1,a^{(2)}_- 
+k_*-1;1-k_1+k_2; z\right].\label{eq:fZsol2}
\end{align}
Here, $C_1$ and $C_2$ are normalising constants that ensure the integral 
constraints 
\begin{align*}
&\int_0^1 f_*(z)\D z=\pi_* \\
& \int_0^1 f_Z(z)\D z = \int_0^1 \frac{1}{z} f_*(z)\D z = 1
\end{align*} 
are satisfied. These constants are
conveniently obtained from a Mellin transform identity (see 
Ref.~\cite{Slater}, p. 152):
\begin{align*}
 C_1 &= 
\frac{\Gamma(k_2-k_1)}{\Gamma(k_1)\Gamma(k_2)}\,
\frac{\Gamma\left(1+k_1-a^{(1)}_+\right)}{\Gamma\left(1-a^{(1)}_+\right)}\,
\frac{\Gamma\left(1+k_1-a^{(1)}_-\right)}{\Gamma\left(1-a^{(1)}_-\right)}; \\
 C_2 &= 
\frac{\Gamma(k_1-k_2)}{\Gamma(k_1)\Gamma(k_2)}\,
\frac{\Gamma\left(1+k_2-a^{(2)}_+\right)}{\Gamma\left(1-a^{(2)}_+\right)}\,
\frac{\Gamma\left(1+k_2-a^{(2)}_-\right)}{\Gamma\left(1-a^{(2)}_-\right)},
\end{align*}
where $\Gamma(.)$ is the Gamma function~\cite{Abramowitz}.
Eq.~\eqref{eq:fZsol2} 
is useful for comparisons with the 2-state random telegraph model.

\end{document}